\documentclass[10pt,twocolumn]{article}
\usepackage[T1]{fontenc}
\usepackage[utf8]{inputenc}
\usepackage[english]{babel}
\usepackage[a4paper,margin=18mm,columnsep=7mm]{geometry}
\usepackage{newtxtext}
\usepackage{amsmath,amssymb,amsfonts}
\usepackage{graphicx}
\usepackage{subcaption}
\usepackage{microtype}
\usepackage{float}
\usepackage{placeins}
\usepackage{cite}
\usepackage[hidelinks,pdfencoding=auto]{hyperref}
\usepackage{bbm}
\numberwithin{equation}{section}
\numberwithin{figure}{section}
\newcommand{\Int}{\int\limits}
\newsavebox{\equationbox}
\newcommand{\mathfit}[2]{%
  \sbox{\equationbox}{$\displaystyle #2$}%
  \typeout{MATHWIDTH:#1:\the\wd\equationbox:LIMIT:\the\linewidth}%
  \ifdim\wd\equationbox>\dimexpr\linewidth-20pt\relax
    \resizebox{\dimexpr\linewidth-20pt\relax}{!}{\usebox{\equationbox}}%
  \else\usebox{\equationbox}\fi}
\hypersetup{pdftitle={Finite Systems in the Space of Projective Measurements},
 pdfauthor={Miguel Flores Rodriguez}}
\usepackage{authblk}

\title{Finite Systems in the Space of Projective Measurements}
\author[1]{Miguel Flores R.}
\author[1]{Carlos A. Muñoz}

\affil[1]{Departamento de Física,
Centro Universitario de Ciencias Exactas e Ingeniería,\\
Universidad de Guadalajara, Revolución 1500,
44410 Guadalajara, México}

\date{}
\begin{document}
\raggedbottom
\twocolumn[\begin{@twocolumnfalse}
\maketitle
\vspace{-5mm}
\begin{abstract}
We investigate the dynamics of two-level systems using the quasidistribution $Q$ function projected onto the space of collective measurements. Particular attention is given to evolution under Hamiltonians that are invariant under particle permutations, for pure states that are either invariant or noninvariant under such permutations. This requires constructing an approximate analog of the star product between quasidistribution functions: the projected $Q$ function of a product of two operators is expressed approximately in terms of the individual operators' $Q$ and $P$ functions. We rewrite the Liouville equation to determine approximate state dynamics and compare the result with exact evolution using the Hellinger distance. Overall, the method performs well for permutation-invariant pure states and Hamiltonians, whereas deviations arise for arbitrary pure states and noninvariant Hamiltonians.

\end{abstract}
\vspace{3mm}
\noindent\small\textbf{Keywords:} discrete phase space; collective measurements; Husimi $Q$ function; projected star product; permutation symmetry; quantum dynamics.
\par\vspace{5mm}
\end{@twocolumnfalse}]
\section{Introduction}
The considerable interest in describing quantum systems containing many particles, that is, systems approaching the macroscopic limit, has motivated different methods for describing their physical properties under Hamiltonian evolution \cite{1,2}. However, these systems are difficult to analyze both theoretically and experimentally. The complexity of representing a system in the macroscopic limit grows exponentially with particle number, and only some of its properties can be meaningfully characterized. The experimental situation is similar: measurements of individual particles' quantum properties do not provide complete information about the state of the entire system. Furthermore, macroscopic systems often contain indistinguishable particles, making their characterization more difficult.

A common discrete representation maps quantum states onto a $2^N\times 2^N$ grid, known as discrete phase space \cite{DPSFF,intro2}. Its disadvantage is that it is useful mainly for a small number of particles. At larger particle numbers, plots become difficult to interpret because of the quantity of information represented. Even for some states with relatively few particles, this representation does not clearly convey the density of information or its location in phase space, complicating the analysis of Hamiltonian evolution \cite{Qmnk,SymmetricDCS,CIS}.

In response to these difficulties, this work studies many-particle quantum dynamics using discrete quasidistributions, specifically the $Q$ and $P$ functions \cite{Andrei}. Its central task is to construct an analog of the star product for the $Q$ function, a product defined in discrete phase space that gives the $Q$ function of a product of operators \cite{Andrei,tesis}. This enables the Liouville equation to be rewritten in terms of distribution functions represented in the space of collective measurements and provides a method for analyzing quantum evolution.

The space of collective measurements offers an approach to characterizing quantum systems in the macroscopic limit \cite{Qmnk}. A discrete-phase-space $Q$ function can be mapped into this space while retaining the information required to study global properties. The resulting function is more convenient for visualization and for analyzing a given quantum state \cite{Andrei}. This approach addresses the difficulty of measuring specific particles, which is generally impossible in the setting considered here, by focusing on information about the system's global properties.

Accordingly, state evolution in the space of collective measurements is expected to continue providing useful system information while revealing the dynamics generated by a Hamiltonian.

We therefore propose a mapping of the $Q$ function for a product of two operators into the space of collective measurements. Here the operators are a density matrix $\hat{\rho}$ and a Hamiltonian $\hat{H}$, and both permutation-invariant and noninvariant quantum states are studied.

The analysis also considers Hamiltonians that are either invariant or noninvariant under particle permutations.

Finally, the Liouville equation is rewritten in the proposed representation to analyze the dynamics of the system's quantum states.

\section{Discrete Phase Space and Collective Measurements}
Several methods have been proposed for representing $N$-particle quantum systems. We focus on collections of two-level systems, such as $N$-qubit systems, spin chains, and electrons in a magnetic solid. For this purpose, one can construct a $2^N\times 2^N$ grid called discrete phase space \cite{7,8}.

Discrete functions on this grid can be associated directly with quantum states: a density matrix is assigned a function defined on discrete phase space. Likewise, Hermitian operators acting in Hilbert space can be assigned phase-space functions. Common examples are the Wigner function, the Glauber $P$ function, and the Husimi $Q$ function. This work focuses on the latter two.

\subsection{Discrete space}
For $N$ two-level particles, the Hilbert space $\mathcal{H}_{2^N}=\mathcal{H}_2^{\otimes N}$ has dimension $2^N$. Its states can be labeled using elements of the finite field $\mathbb{F}_{2^N}$, regarded as a linear space with expansions in an abstract basis $\lbrace\theta_1,\ldots,\theta_N\rbrace$. Each $\alpha\in\mathbb{F}_{2^N}$ can thus be written as

\begin{equation}
\mathfit{1}{\begin{aligned}
& \alpha= \sum_{i=1}^{N} a_i \theta_i,   a_i \in  \mathbb{Z}_2,
\end{aligned}}\label{eq:s2_1}
\end{equation}
This associates field elements with binary strings, $\alpha\Leftrightarrow(a_1,a_2,\ldots,a_N)$.

The basis $\lbrace\theta_1,\ldots,\theta_N\rbrace$ can be chosen to be self-dual, so that $tr(\theta_i\theta_j)=\delta_{i,j}$, with $tr(\alpha)=\alpha+\alpha^2+\cdots+\alpha^{2^{N-1}}$. Consequently, $a_i=tr(\alpha\theta_i)$.\\

We label the Hilbert-space states of an $N$-qubit system by finite-field elements through $\vert\kappa\rangle=\vert k_1,\ldots,k_N\rangle$, where $k_i\in\mathbb{Z}_2$. We also define a $2^N\times 2^N$ grid, henceforth called discrete phase space. Its coordinates $(\alpha,\beta)$ are binary $N$-component strings, $\alpha=(a_1,\ldots,a_N)$ and $\beta=(b_1,\ldots,b_N)$, with $a_i,b_i\in\mathbb{Z}_2$. Define the operator basis $Z_\alpha,X_\beta$ \cite{Qmnk} by

\begin{equation}
\mathfit{2}{\begin{aligned}
& Z_{\alpha}=\sigma_{z}^{a_1} \otimes \cdots \otimes \sigma_{z}^{a_N},   X_{\beta}=\sigma_{x}^{b_1} \otimes \cdots \otimes \sigma_{x}^{b_N},
\end{aligned}}\label{eq:s2_2}
\end{equation}
Here $\sigma_z=\vert0\rangle\langle0\vert-\vert1\rangle\langle1\vert$ and $\sigma_x=\vert0\rangle\langle1\vert+\vert1\rangle\langle0\vert$ are Pauli matrices. The generalized Pauli operators $Z_\alpha$ and $X_\beta$ obey

\begin{equation}
\mathfit{3}{\begin{aligned}
& Z_{\alpha} X_{\beta}= (-1)^{tr(\alpha \beta)}  X_{\beta}Z_{\alpha}.
\end{aligned}}\label{eq:s2_3}
\end{equation}
Discrete coherent-type states, labeled by the phase-space coordinates $(\alpha,\beta)$ \cite{SymmetricDCS}, are defined by

\begin{equation}
\mathfit{4}{\begin{aligned}
& \vert \alpha, \beta \rangle =Z_{\alpha} X_{\beta} \vert \xi \rangle,
\end{aligned}}\label{eq:s2_4}
\end{equation}
Because they form a complete set over discrete phase space, they satisfy

 \begin{equation}
\mathfit{5}{\begin{aligned}
& \sum_{\alpha,\beta} \vert \alpha, \beta \rangle  \langle \alpha, \beta \vert = 2^N \mathbbm{1},
\end{aligned}}\label{eq:s2_5}
\end{equation}
where the sum runs over all strings $\alpha,\beta$. The fiducial state $\vert\xi\rangle$ is an $SU(2)$ coherent state that is invariant under particle permutations. It can therefore be written as a product of $N$ identical two-level states:

\begin{equation}
\mathfit{6}{\begin{aligned}
& \vert \xi \rangle \propto \vert \xi \rangle_1 \otimes \vert \xi \rangle_2 \otimes \cdots \otimes \vert \xi \rangle_N.
\end{aligned}}\label{eq:s2_6}
\end{equation}
In the computational basis, $\vert\xi\rangle$ is

 \begin{equation}
\mathfit{7}{\begin{aligned}
& \vert \xi \rangle = \frac{1}{(1+\vert \xi \vert^2 )^{N/2}} \sum_{k} \xi^{h(k)} \vert k \rangle,
\end{aligned}}\label{eq:s2_7}
\end{equation}
Here $h(k)=\sum_i k_i$ is the Hamming weight, namely the number of nonzero coefficients when $k$ is expanded in the self-dual basis. The parameter $\xi$ may take complex values; a choice that exploits symmetry in the $P$ function \cite{CIS} is

\begin{equation}
\mathfit{8}{\begin{aligned}
& \xi = \frac{\sqrt{3}-1}{\sqrt{2}} e^{\pi i /4}.
\end{aligned}}\label{eq:s2_8}
\end{equation}
We use this value to define the discrete coherent state from which the $Q$ and $P$ functions are constructed.

\subsection{Distribution functions in discrete space}
We restrict attention to the $Q$ and $P$ functions because analyzing dynamics with the Wigner function involves an arbitrary choice of the phase accompanying the displacement operator $Z_\alpha X_\beta$ \cite{ref1,ref2,Gaeta,2017}.

\subsubsection{The \texorpdfstring{$P$}{P} function}
For an arbitrary operator $\hat{A}$, the discrete $P$ function is defined through its expansion in coherent states:

\begin{equation}
\mathfit{9}{\begin{aligned}
& \hat{A}= \sum_{\alpha, \beta} P_{A}(\alpha,\beta) \vert \alpha, \beta \rangle \langle \alpha, \beta \vert,
\end{aligned}}\label{eq:s2_9}
\end{equation}
It can also be defined using a kernel:

\begin{equation}
\mathfit{10}{\begin{aligned}
& P_{A}(\alpha,\beta)=tr[ \hat{A} \Delta_{P}(\alpha,\beta)],
\end{aligned}}\label{eq:s2_10}
\end{equation}
The mapping kernel $\Delta_P(\alpha,\beta)$ is

\begin{equation}
\mathfit{11}{\begin{aligned}
& \Delta_{P}(\alpha,\beta)=\frac{1}{2^{2N}}\sum_{\mu, \gamma} (-1)^{\alpha \gamma + \beta \mu} [\langle \xi \vert Z_{\mu} X_{\gamma} \vert \xi \rangle ]^{-1} Z_{\mu} X_{\gamma}
\end{aligned}}\label{eq:s2_11}
\end{equation}
where $\alpha\gamma=\sum_i\alpha_i\gamma_i$.

\subsubsection{The \texorpdfstring{$Q$}{Q} function}
\begin{figure}[H]
\centering
\begin{subfigure}[t]{.98\linewidth}
\centering
\includegraphics[width=\linewidth]{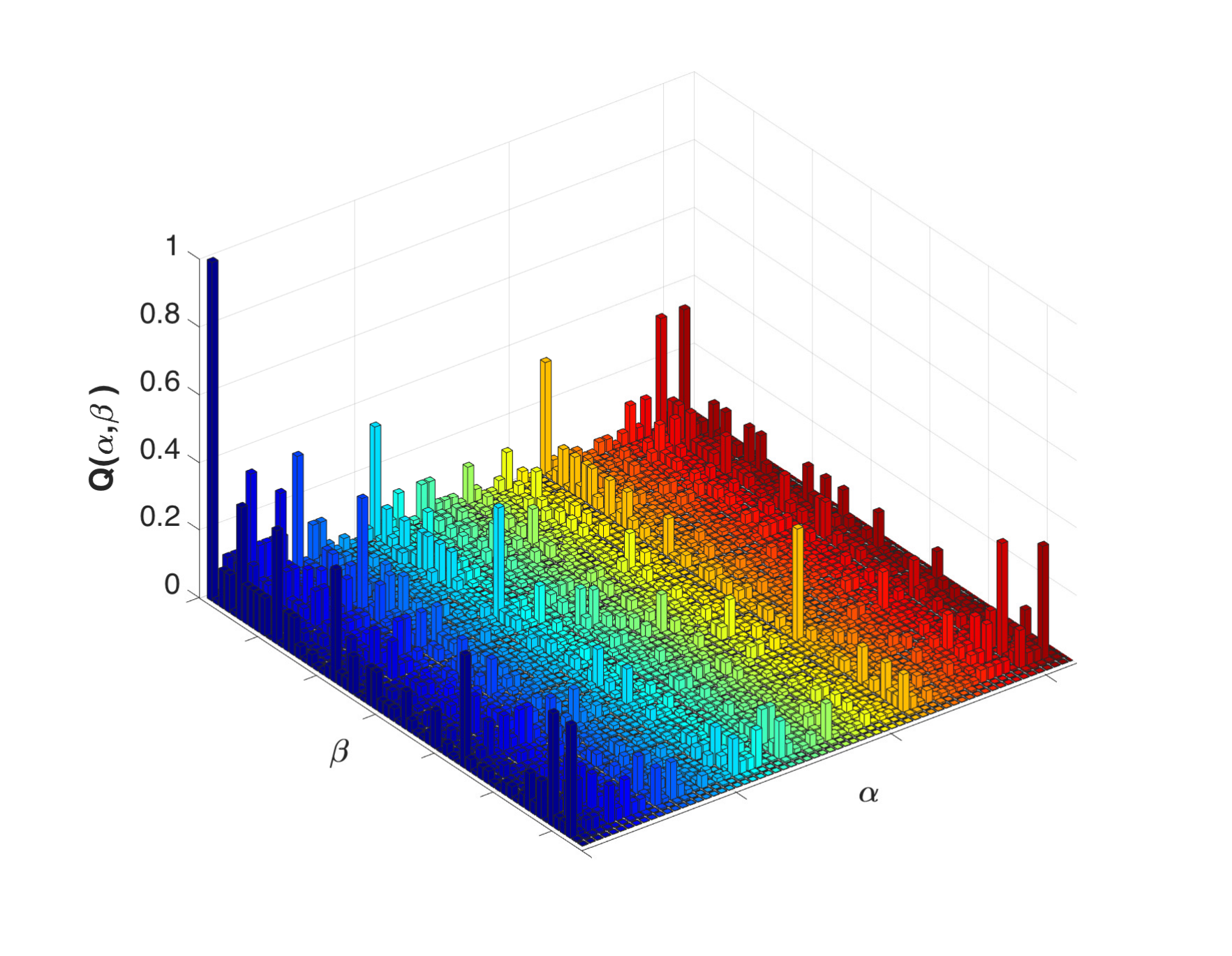}
\caption{}
\end{subfigure}\par\medskip
\begin{subfigure}[t]{.98\linewidth}
\centering
\includegraphics[width=\linewidth]{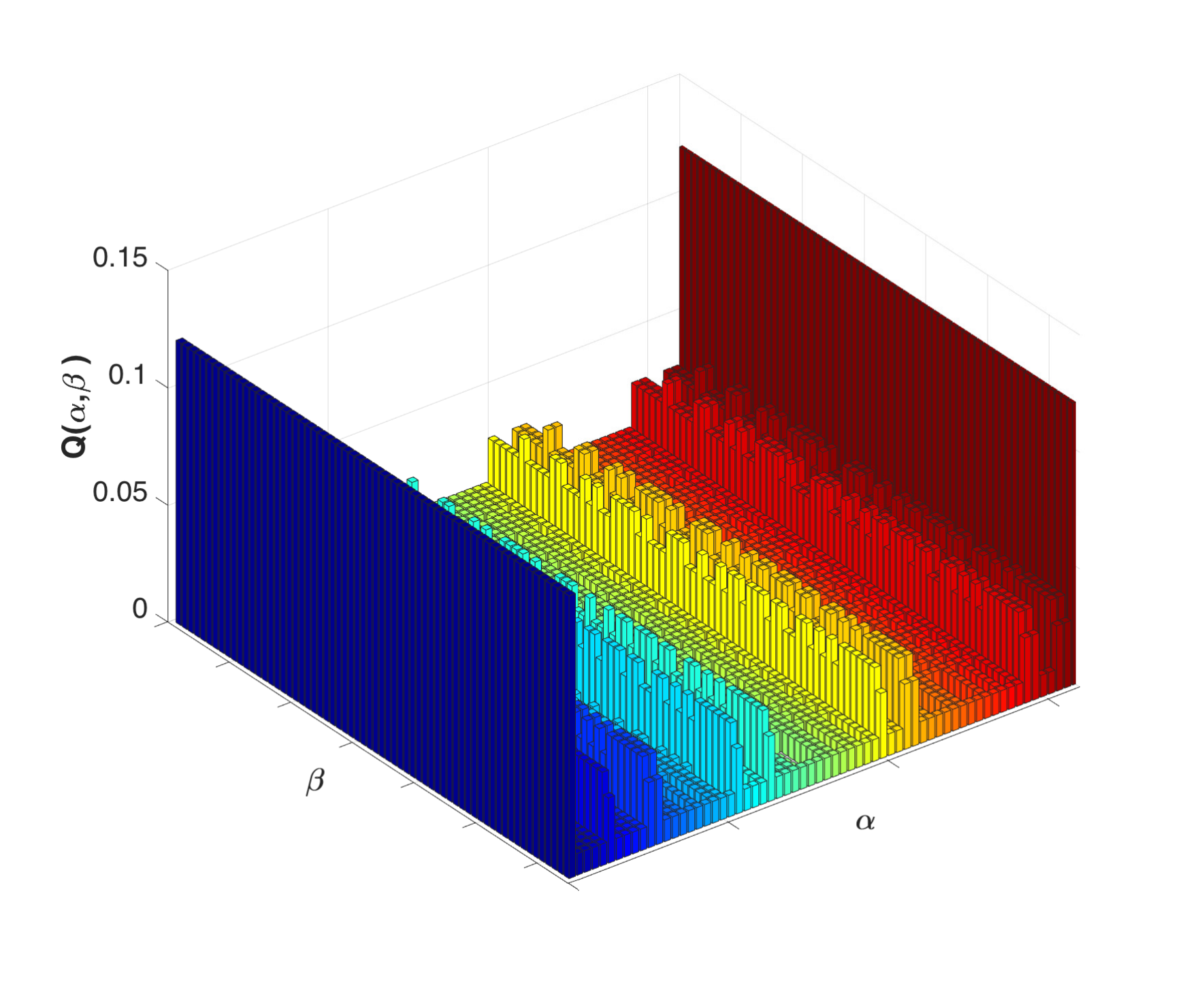}
\caption{}
\end{subfigure}
\caption{Discrete $Q$ functions for (a) the coherent state $\vert\xi\rangle$ in Eq.~\eqref{eq:s2_7} and (b) the state $\vert GHZ\rangle=[\vert0\cdots0\rangle+\vert1\cdots1\rangle]/\sqrt{2}$.}
\label{fig:s2_1}
\end{figure}

The discrete $Q$ function is defined generally as the expectation value

\begin{equation}
\mathfit{12}{\begin{aligned}
& Q_{A}=\langle \alpha, \beta \vert \hat{A} \vert \alpha, \beta \rangle,
\end{aligned}}\label{eq:s2_12}
\end{equation}
As with the $P$ function, it can be constructed using a kernel:

\begin{equation}
\mathfit{13}{\begin{aligned}
& Q_{A}(\alpha,\beta)=tr[ \hat{A} \Delta_{Q}(\alpha,\beta)],
\end{aligned}}\label{eq:s2_13}
\end{equation}
where

\begin{equation}
\mathfit{14}{\begin{aligned}
& \Delta_{Q}(\alpha,\beta)=\frac{1}{2^{2N}}\sum_{\mu, \gamma} (-1)^{\alpha \gamma + \beta \mu} [\langle \xi \vert Z_{\mu} X_{\gamma} \vert \xi \rangle ]Z_{\mu} X_{\gamma} \\
& =\vert \alpha, \beta \rangle \langle \alpha, \beta \vert.
\end{aligned}}\label{eq:s2_14}
\end{equation}
For the state distributions, the normalization is

\begin{equation}
\mathfit{15}{\begin{aligned}
& \sum_{\alpha,\beta}Q(\alpha,\beta)= 2^N,   \sum_{\alpha,\beta}P(\alpha,\beta)=1.
\end{aligned}}\label{eq:s2_15}
\end{equation}
These mappings are dual in the sense that an operator $\hat{A}$ satisfies

\begin{equation}
\mathfit{16}{\begin{aligned}
& \hat{A}=\sum_{\alpha,\beta}Q_{A}(\alpha,\beta)\Delta_{P}(\alpha,\beta),          \hat{A}=\sum_{\alpha,\beta}P_{A} (\alpha,\beta)\Delta_{Q}(\alpha,\beta),
\end{aligned}}\label{eq:s2_16}
\end{equation}
These identities follow from $tr[\Delta_P(\alpha,\beta)\Delta_Q(\gamma,\delta)]=\delta_{\alpha,\gamma}\delta_{\beta,\delta}$. The expectation value of an operator $\hat{A}$ is obtained from

\begin{equation}
\mathfit{17}{\begin{aligned}
& \langle \hat{A} \rangle = \sum_{\alpha,\beta} P_{A}(\alpha,\beta)Q_{\rho}(\alpha, \beta).
\end{aligned}}\label{eq:s2_17}
\end{equation}
In addition, the $Q$ and $P$ functions are related by

\begin{equation*}
\mathfit{18}{\begin{aligned}
& Q(\alpha,\beta)= \sum_{\gamma, \mu} 3^{-(h(\alpha+\gamma)+h(\beta+\mu)+h(\alpha+\gamma+\beta+\mu))/2} \P(\gamma,\mu),
\end{aligned}}
\end{equation*}
\begin{equation*}
\mathfit{19}{\begin{aligned}
& P(\alpha,\beta)=\frac{1}{2^{2N}}\sum_{\gamma, \mu} (-1)^{\alpha \mu+ \beta \gamma } 3^{(h(\gamma)+h(\mu)+h(\gamma+\mu))/4} \\
& \times Tr[(\rho( (-1)^{\gamma \mu}  Z_{\gamma} X_{\mu})].
\end{aligned}}
\end{equation*}
In particular, the $Q$ function of an $SU(2)$ coherent state is

\begin{equation*}
\mathfit{20}{\begin{aligned}
& Q_{\vert \xi \rangle} (\alpha, \beta)=3^{-(h(\alpha)+h(\beta)+h(\alpha+\beta))/2},
\end{aligned}}
\end{equation*}
For $\vert GHZ\rangle=[\vert0\cdots0\rangle+\vert1\cdots1\rangle]/\sqrt{2}$, the $Q$ function is

\begin{equation*}
\mathfit{21}{\begin{aligned}
& Q_{GHZ} (\alpha, \beta) = \frac{|\xi^{h(\alpha)}+(-1)^{h(\alpha)} \xi^{N-h(\beta)} |^2}{2(1 + | \xi |^2)^N},
\end{aligned}}
\end{equation*}
Fig.~\ref{fig:s2_1} shows these $Q$ functions in discrete phase space.

Fig.~\ref{fig:s2_1} illustrates a disadvantage of studying quantum systems using quasidistributions in discrete space: the display contains an excessive amount of information. This becomes problematic as the particle number increases and collective properties are investigated.

\subsection{Measurements of collective operators}
As discussed in the introduction, this study is principally concerned with discrete quantum systems containing many particles, that is, systems approaching the macroscopic limit.

Studying the dynamics of individual particles is then inefficient and sometimes impossible because the particles cannot be distinguished. Moreover, individual particle states are not needed to describe collective system properties. We therefore consider measurements of collective observables. They provide global information and a useful statistical description of the quantum system. In the experimental setting considered here, such collective measurements are the accessible measurements \cite{opcolectivos}.

For $N$ qubits, the measured observables are polynomial functions of the collective operators

\begin{equation}
\mathfit{22}{\begin{aligned}
& S_{x,y,z}=\sum_{i=1}^{N}\sigma_{x,y,z}^{(i)}
\end{aligned}}\label{eq:s2_18}
\end{equation}
where $\sigma_{x,y,z}^{(i)}$ are Pauli matrices acting on particle $i$.

\subsection{The space of collective measurements}
The three-dimensional space of collective measurements was introduced as a convenient representation and analysis tool for quantum systems in the macroscopic limit. The projected $Q$ function contains complete, nonredundant information about measurements of all moments of any permutation-invariant observable, for an arbitrary state that need not be symmetric. In particular, $N$-particle states can be classified by their localization in measurement space. Localized states occupy a small region and require fewer measured parameters for their characterization than nonlocalized states, which extend throughout the space \cite{Gaeta,2017}.

\subsubsection{States in the space of collective measurements}
The objective is to define projected quasidistributions that describe the collective properties of a quantum system.

\begin{figure}[H]
\centering
\begin{subfigure}[t]{.98\linewidth}
\centering
\includegraphics[width=\linewidth]{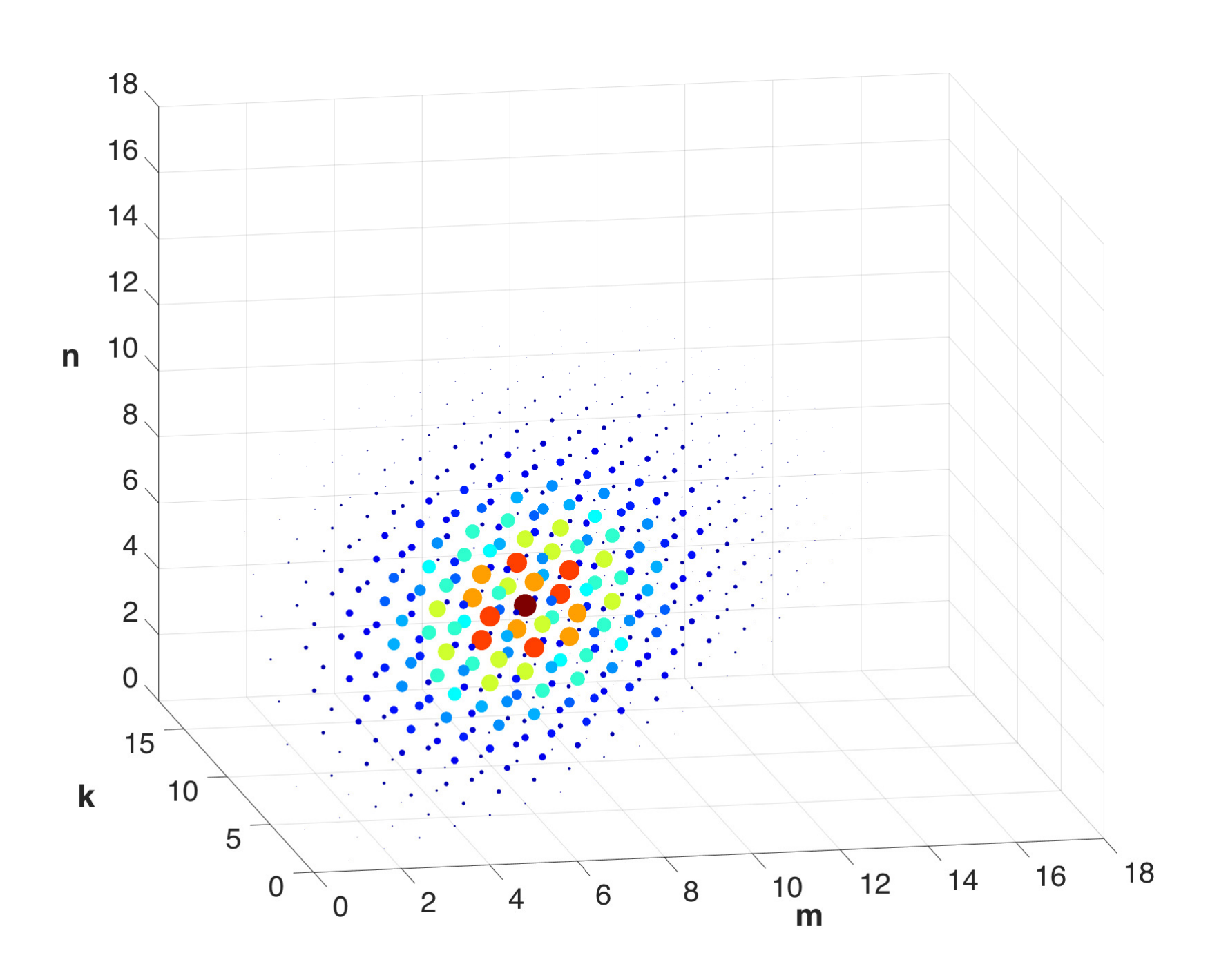}
\caption{}
\end{subfigure}\par\medskip
\begin{subfigure}[t]{.98\linewidth}
\centering
\includegraphics[width=\linewidth]{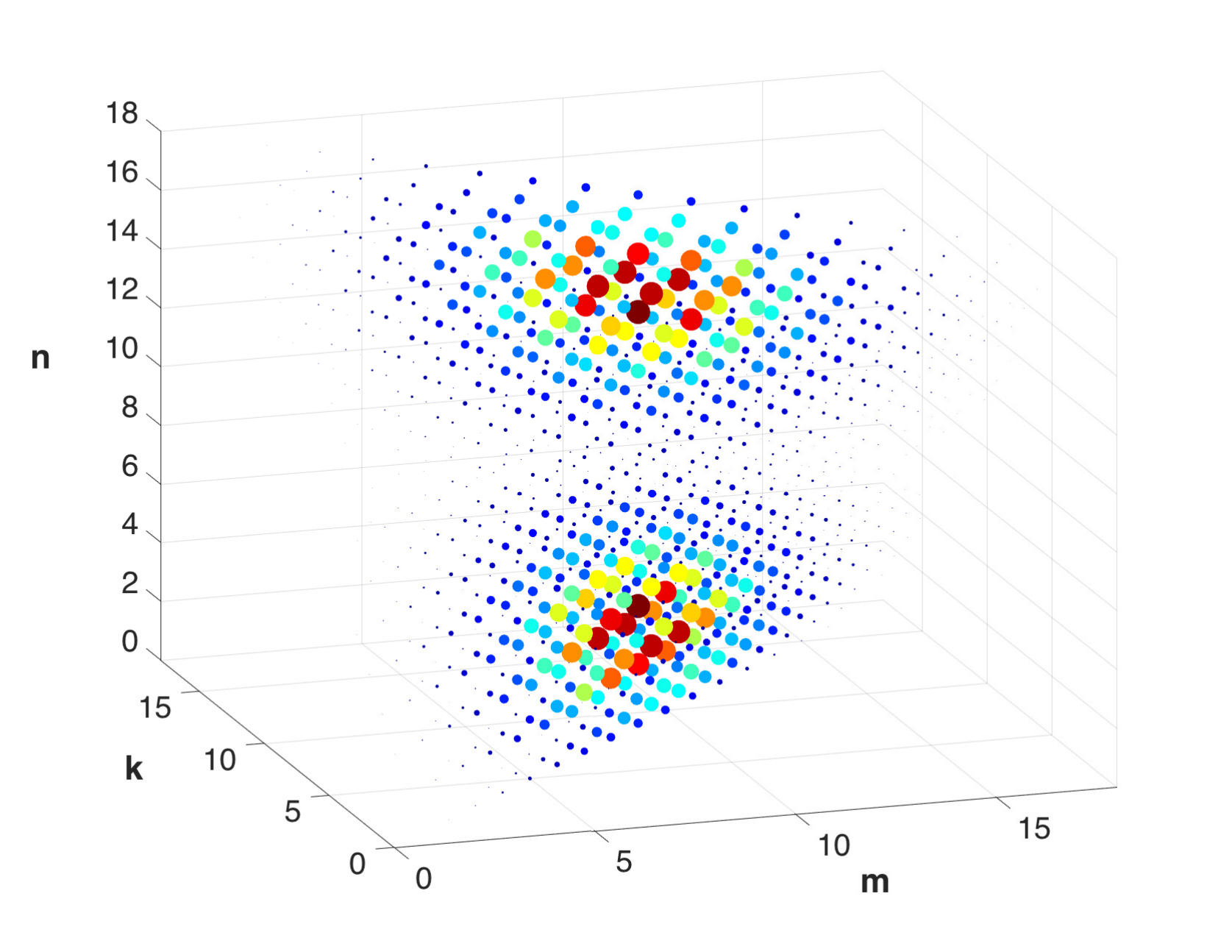}
\caption{}
\end{subfigure}\par\medskip
\begin{subfigure}[t]{.98\linewidth}
\centering
\includegraphics[width=\linewidth]{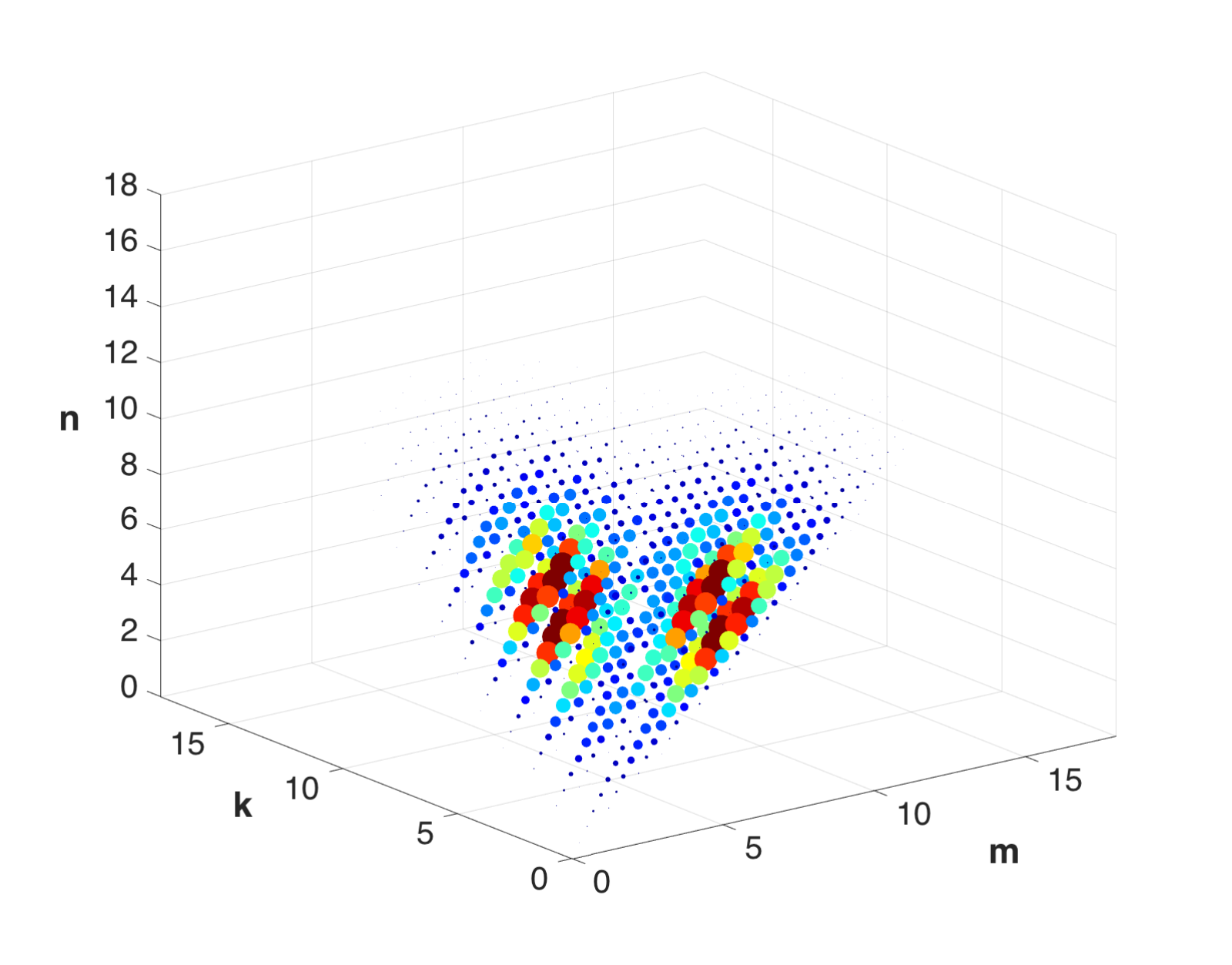}
\caption{}
\end{subfigure}
\caption{Projected $\tilde{Q}(m,n,k)$ functions for $N=18$: (a) the coherent state in Eq.~\eqref{eq:s2_6}; (b) $\vert GHZ\rangle=[\vert0\cdots0\rangle+\vert1\cdots1\rangle]/\sqrt{2}$; and (c) $\vert W\rangle=[\vert100\cdots0\rangle+\vert010\cdots0\rangle+\cdots+\vert00\cdots01\rangle]/\sqrt{N}$.}
\label{fig:s2_2}
\end{figure}

To define the projections of the $Q$ and $P$ functions, we use the fact that for a symmetric collective operator $\hat{A}$, invariant under particle permutations, these functions depend on $(\alpha,\beta)$ only through the Hamming weights of $\alpha$, $\beta$, and $\alpha+\beta$:

\begin{equation}
\mathfit{23}{\begin{aligned}
& P_{A}(\alpha,\beta)=P_{A}(h(\alpha),h(\beta),h(\alpha+\beta)).
\end{aligned}}\label{eq:s2_19}
\end{equation}
For example, the collective operators in Eq.~\eqref{eq:s2_18} yield

\begin{equation*}
\mathfit{24}{\begin{aligned}
& P_{S_x}=\frac{\sqrt{3}}{2^N} [ N - 2h(\alpha)],
\end{aligned}}
\end{equation*}
\begin{equation*}
\mathfit{25}{\begin{aligned}
& P_{S_y}=\frac{\sqrt{3}}{2^N} [ N - 2h(\alpha +\beta)],
\end{aligned}}
\end{equation*}
\begin{equation*}
\mathfit{26}{\begin{aligned}
& P_{S_z}=\frac{\sqrt{3}}{2^N} [ N - 2h(\beta)].
\end{aligned}}
\end{equation*}
Expressing a collective operator's expectation value using Eq.~\eqref{eq:s2_17} allows the projected $Q$ function to be introduced through

\begin{equation}
\mathfit{27}{\begin{aligned}
& \langle \hat{A} \rangle = \sum_{m,n}\sum_{k} P_{A}(m,n,k)\tilde{Q}_{\rho}(m,n,k),
\end{aligned}}\label{eq:s2_20}
\end{equation}
Here $m=h(\alpha)$, $n=h(\beta)$, and $k=h(\alpha+\beta)$. The sums run over $m,n=0,\ldots,N$, while $k$ takes steps of two: $k=\lvert m-n\rvert,\lvert m-n\rvert+2,\ldots,\min(m+n,N,2N-m-n)$. These constraints arise from the possible particle permutations. The function $\tilde{Q}$ is the projection of $Q$ onto the three-dimensional space of collective measurements:

\begin{equation}
\mathfit{28}{\begin{aligned}
& \tilde{Q}_{\rho}(m,n,k)=\sum_{\alpha,\beta} Q_{\rho}(\alpha,\beta) \delta_{m,h(\alpha)} \delta_{n,h(\beta)} \delta_{k,h(\alpha+\beta)}.
\end{aligned}}\label{eq:s2_21}
\end{equation}
For a permutation-invariant density operator $\hat{\rho}$, Eq.~\eqref{eq:s2_19} also allows us to write

\begin{equation}
\mathfit{29}{\begin{aligned}
& \tilde{Q}_{\rho}(m,n,k)=Q_{\rho}(m,n,k) R_{m,n,k},
\end{aligned}}\label{eq:s2_22}
\end{equation}
where the multiplicity $R_{m,n,k}$ is

\begin{equation*}
\mathfit{30}{\begin{aligned}
& R_{m,n,k}=\frac{N!}{(\frac{m+n-k}{2})!(\frac{2N-m-n-k}{2})!(\frac{n-m+k}{2})!(\frac{m-n+k}{2})!}.
\end{aligned}}
\end{equation*}
Similarly, for a permutation-invariant Hamiltonian, the projected $P$ function is

\begin{equation}
\mathfit{31}{\begin{aligned}
& \tilde{P}_{H}(m,n,k)=P_{H}(m,n,k) R_{m,n,k}.
\end{aligned}}\label{eq:s2_23}
\end{equation}
The function $\tilde{Q}$ in Eq.~\eqref{eq:s2_21} is defined on a discrete three-dimensional tetrahedron with vertices

\begin{equation*}
\mathfit{32}{\begin{aligned}
& \lbrace(0,0,0),(0,N,N),(N,0,N),(N,N,0)\rbrace.
\end{aligned}}
\end{equation*} The projected $\tilde{Q}$ function contains all the information needed to obtain the expectation value of any permutation-invariant observable using Eq.~\eqref{eq:s2_20}. It has been used to classify states by localization in measurement space \cite{Qmnk}. Fig.~\ref{fig:s2_2} shows examples of $\tilde{Q}(m,n,k)$. Each sphere denotes a point inside the tetrahedron; its size and color are proportional to the value of $\tilde{Q}$, with red indicating larger values and blue smaller values. Compared with the discrete-grid functions in Fig.~\ref{fig:s2_1}, these projected functions provide a clearer visualization of where the information density is concentrated. They can also be used to analyze global system properties in the asymptotic limit \cite{Qmnk,Gaeta}. However, projection discards part of the original state information retained by the full phase-space $Q$ function. The projected function retains the information required for expectation values of collective operators through Eq.~\eqref{eq:s2_20}.\\

\subsection{Quantum states in two-level systems}
We analyze two types of states: those invariant under particle permutations and those that are not. In the first case, exchanging subsystems leaves the state unchanged; in the second, it changes the state. Both types are evolved using permutation-invariant and noninvariant Hamiltonians.

\subsubsection{Permutation-invariant states}
We begin with the GHZ (Greenberger-Horne-Zeilinger) state \cite{GHZ}:

\begin{equation}
\mathfit{33}{\begin{aligned}
& \vert GHZ \rangle =[\vert 0\cdots0 \rangle + \vert 1\cdots1 \rangle] /\sqrt{2},
\end{aligned}}\label{eq:s2_24}
\end{equation}

This entangled state is a superposition of all particles occupying the ground state $\vert0\rangle$ and all particles occupying the excited state $\vert1\rangle$.

Another permutation-invariant state is $\vert W\rangle$:

\begin{equation}
\mathfit{34}{\begin{aligned}
& \vert W \rangle =\frac{\vert 100 \cdots 0\rangle + \vert 010 \cdots 0\rangle + \cdots \vert 00 \cdots 01 \rangle}{\sqrt{N}},
\end{aligned}}\label{eq:s2_25}
\end{equation}
It is an equal-coefficient superposition of all configurations with exactly one particle in $\vert1\rangle$ and all others in $\vert0\rangle$. It is the $K=1$ special case of a Dicke state.

For a three-particle system, both GHZ and W states are nonbiseparable. They nevertheless differ importantly: after a W state loses one particle, the remaining particles are still entangled, whereas a GHZ state becomes separable after one particle is lost.

Both states are used in quantum information, for example in cryptographic protocols involving two-level systems \cite{W}.

We also analyze the coherent state of Eq.~\eqref{eq:s2_7}:

\begin{equation}
\mathfit{35}{\begin{aligned}
& \vert \xi \rangle = \frac{1}{(1+\vert \xi \vert^2 )^{N/2}} \sum_{k} \xi^{h(k)} \vert k \rangle,  \xi = \frac{\sqrt{3}-1}{\sqrt{2}} e^{\pi i /4},
\end{aligned}}\label{eq:s2_26}
\end{equation}
It is permutation-invariant because, as in Eq.~\eqref{eq:s2_6}, it consists of identical component states.

Finally, we study the Dicke state $\vert k=N/2,N\rangle$ \cite{CIS}. The general definition is

\begin{equation}
\mathfit{36}{\begin{aligned}
& \vert k, N \rangle = \sqrt{\frac{k!(N-k)!}{N!}} \sum_p \mathcal{P}_k (\vert 1 \rangle_1 ... \vert 1 \rangle_k \vert 0 \rangle_{k+1} ... \vert 1 \rangle_N),
\end{aligned}}\label{eq:s2_27}
\end{equation}
where $\lbrace\mathcal{P}_k\rbrace$ denotes all possible particle permutations.

\subsubsection{States that are not permutation-invariant}
The first noninvariant state is the domain-wall state,

\begin{equation}
\mathfit{37}{\begin{aligned}
& \vert DW \rangle =\frac{\vert 0 \ldots0 \rangle + \vert 10 \ldots 0 \rangle+\vert 110 \ldots 0 \rangle + \ldots + \vert 1 \ldots 1 \rangle}{\sqrt{N+1}},
\end{aligned}}\label{eq:s2_28}
\end{equation}
It describes a system whose configurations interpolate from particles in the ground state $\vert0\rangle$ at one end to particles in the excited state $\vert1\rangle$ along the chain. An interpretation arises in magnetism: magnetic moments point in one direction at one end, gradually reorient along the chain, and point in the opposite direction at the other end \cite{DWall}.

Another state in this category is the NOON state,

\begin{equation}
\mathfit{38}{\begin{aligned}
& \vert NOON \rangle =\frac{\vert 0 \ldots 0, 1, \ldots 1 \rangle +\vert 1 \ldots 1, 0, \ldots 0 \rangle  }{\sqrt{2}},
\end{aligned}}\label{eq:s2_29}
\end{equation}
This is a superposition of configurations with $N/2$ particles in $\vert0\rangle$ and $N/2$ in $\vert1\rangle$. In optics, NOON states describe a superposition of $N/2$ photons in mode $a$ with none in mode $b$, and $N/2$ photons in mode $b$ with none in mode $a$ \cite{NOON}.

We also study the singlet-type state

\begin{equation}
\mathfit{39}{\begin{aligned}
& \vert S\rangle = \bigotimes_{i=1,3,...,N/2+1} \frac{\vert 0 \rangle_i \vert 1 \rangle_{i+1}-\vert 1 \rangle_i \vert 0 \rangle_{i+1} }{\sqrt{2}},
\end{aligned}}\label{eq:s2_30}
\end{equation}
It describes entangled particles with zero total angular momentum. The state has $N$ particles, half with spins in one direction and half in the opposite direction. A physical example of a singlet is positronium, consisting of an electron and a positron bound by their opposite charges \cite{S}.

\subsection{System evolution}
The central objective is to study quantum dynamics in measurement space $(m,n,k)$. We start from the Liouville equation,

\begin{equation}
\mathfit{40}{\begin{aligned}
& \mathit{i} \hbar \frac{d}{dt} \hat{\rho}(t) =[\hat{\rho(t)},\hat{H}],
\end{aligned}}\label{eq:s2_31}
\end{equation}
and rewrite it in terms of a $Q$ function as \cite{tesis}

\begin{equation}
\mathfit{41}{\begin{aligned}
& \mathit{i}\hbar  \frac{d}{dt} \mathit{Q}_{\rho} =Q_{[\hat{H},\hat{\rho}]} =\mathit{Q}_{H \rho} - \mathit{Q}_{\rho H}.
\end{aligned}}\label{eq:s2_32}
\end{equation}
In discrete phase space, the star product \cite{Andrei,tesis} defines the $Q$ function of a product of two operators:

\begin{equation}
\mathfit{42}{\begin{aligned}
& Q_A \ast Q_B =Q_{AB}= \hat{L}(Q_A,Q_B),
\end{aligned}}\label{eq:s2_33}
\end{equation}
Here $\hat{L}$ is a difference or summation operator acting on a product of $Q$ distributions.

We seek an analog of Eq.~\eqref{eq:s2_33} using $Q$ functions projected onto measurement space $(m,n,k)$.

\section{Methodology}
\subsection{Evolution in the space of collective measurements}
To define the $Q$ function of a product of operators, here the density matrix $\hat{\rho}$ and Hamiltonian $\hat{H}$, we first expand them using Eq.~\eqref{eq:s2_16}:

\begin{equation}
\mathfit{43}{\begin{aligned}
& \hat{\rho}=\sum_{\alpha,\beta} Q_{\rho}(\alpha,\beta) \Delta_P(\alpha,\beta),  \hat{H}=\sum_{\alpha,\beta} P_H(\alpha,\beta) \Delta_Q(\alpha,\beta).
\end{aligned}}\label{eq:s3_1}
\end{equation}
Substituting these expressions into Eq.~\eqref{eq:s2_12} gives

\begin{equation}
\mathfit{44}{\begin{aligned}
& Q_{\rho H}(\mu, \nu)=\langle \mu, \nu \vert [ \sum_{\alpha,\beta} Q_{\rho}(\alpha,\beta) \Delta_P(\alpha,\beta) ] \\
& \times [ \sum_{\alpha',\beta'} P_H(\alpha',\beta') \Delta_Q(\alpha',\beta') ] \vert \mu, \nu \rangle,
\end{aligned}}\label{eq:s3_2}
\end{equation}

\begin{equation*}
\mathfit{45}{\begin{aligned}
& = \sum_{\alpha,\beta} \sum_{\alpha',\beta'} Q_{\rho}(\alpha,\beta) P_H(\alpha',\beta')\\
& \times \langle \mu, \nu  \vert  \Delta_P(\alpha,\beta) \vert \alpha', \beta' \rangle \langle \alpha', \beta'  \vert \mu, \nu \rangle,
\end{aligned}}
\end{equation*}
\begin{equation*}
\mathfit{46}{\begin{aligned}
& = \sum_{\alpha,\beta} \sum_{\alpha',\beta'} Q_{\rho}(\alpha,\beta) P_H(\alpha',\beta')
\Phi_{\alpha, \beta, \alpha', \beta', \mu, \nu},
\end{aligned}}
\end{equation*}
Here
\[
\begin{aligned}
\Phi_{\alpha,\beta,\alpha',\beta',\mu,\nu}
&=\langle\mu,\nu\vert\Delta_P(\alpha,\beta)\vert\alpha',\beta'\rangle \quad\times\langle\alpha',\beta'\vert\mu,\nu\rangle.
\end{aligned}
\]
According to Eq.~\eqref{eq:s2_21}, the projection of $Q$ can be written as

\begin{equation*}
\mathfit{47}{\begin{aligned}
& \tilde{Q}_{\rho H}(p,q,r)= \sum_{\alpha,\beta} \sum_{\alpha',\beta'} Q_{\rho}(\alpha,\beta) P_H(\alpha',\beta') \\
& \times \sum_{\mu,\nu}\Phi_{\alpha, \beta, \alpha', \beta', \mu, \nu}  \delta_{h(\alpha),m}\delta_{h(\beta),n}\delta_{h(\alpha+\beta),k}\delta_{h(\alpha'),m'}
\end{aligned}}
\end{equation*}
\begin{equation}
\mathfit{48}{\begin{aligned}
& \times \delta_{h(\beta'),n'}\delta_{h(\alpha'+\beta'),k'} \delta_{h(\mu),p}\delta_{h(\gamma),q}\delta_{h(\mu+\gamma),r}
\end{aligned}}\label{eq:s3_3}
\end{equation}
At this point, we introduce the approximation

\begin{equation}
\mathfit{49}{\begin{aligned}
& Q_{\rho}(\alpha,\beta) \approx R_{mnk}^{-1} \tilde{Q}_{\rho}(m,n,k),
\end{aligned}}\label{eq:s3_4}
\end{equation}
By Eq.~\eqref{eq:s2_22}, this relation is exact for permutation-invariant states. We expect it to perform well for permutation-invariant Hamiltonians and to yield deviations for noninvariant Hamiltonians, for the reasons discussed in Sec.~2.4. For permutation-invariant Hamiltonians, we therefore propose rewriting Eq.~\eqref{eq:s3_3} as

\begin{equation}
\mathfit{50}{\begin{aligned}
& \tilde{Q}_{\rho H}(p,q,r) \approx \sum_{\substack{m,n,k \\ m',n'k'}} R_{m,n,k}^{-1}  \tilde{Q}_{\rho}(m,n,k) P_H(m',n',k')\\
& \times \Psi^{m,n,k,m',n',k'}_{p,q,r},
\end{aligned}}\label{eq:s3_5}
\end{equation}
where

\begin{equation}
\mathfit{51}{\begin{aligned}
& \Psi^{m,n,k,m',n',k'}_{p,q,r} =  \sum_{\substack{\alpha,\beta \\ \alpha, \beta \\ \mu, \nu}} \langle \mu, \nu \vert  \Delta_P(\alpha,\beta) \vert \alpha', \beta' \rangle \\
& \times \langle \alpha', \beta'  \vert \mu, \nu \rangle \delta_{h(\alpha),m}\delta_{h(\beta),n}\delta_{h(\alpha+\beta),k} \delta_{h(\alpha'),m'}\\
& \times \delta_{h(\beta'),n'}\delta_{h(\alpha'+\beta'),k'} \delta_{h(\mu),p} \delta_{h(\gamma),q}\delta_{h(\mu+\gamma),r}
\end{aligned}}\label{eq:s3_6}
\end{equation}
To simplify $\Psi^{m,n,k,m',n',k'}_{p,q,r}$, replace the Kronecker deltas by their integral representations, $\delta_{m,n}=(2\pi)^{-1}\oint e^{i(m-n)\theta}\,d\theta$. Equation~\eqref{eq:s3_6} becomes

\begin{equation*}
\mathfit{52}{\begin{aligned}
& \Psi^{m,n,k,m',n',k'}_{p,q,r}=\frac{1}{2^{2N}}\frac{1}{(2 \pi)^{9} } \oint e^{\substack{-\mathit{i}xm - \mathit{i}zn-\mathit{i}yk - \mathit{i}x'm' - \mathit{i}z'n'\\ - \mathit{i}y'k' - \mathit{i}up - \mathit{i}wk - \mathit{i}vq}}\\
& \times \sum_{\substack{\alpha_i, \beta_i, \alpha_i', \beta_i' \\ \mu_i, \gamma_i, \nu_i, \eta_i}} e^{\substack{\mathit{i}x \alpha_i+\mathit{i}z \beta_i \mathit{i}y (\alpha_i +\beta_i -2\alpha_i \beta_i) +\mathit{i}x' \alpha_i' +i w \gamma_i  \\+\mathit{i}z' \beta_i'\mathit{i}y'(\alpha_i'+\beta_i' -2\alpha_i'\beta_i')\mathit{i}u \mu_i + \mathit{i}v(\mu_i +\gamma_i -2\mu_i\gamma_i)}}\\
& \times (3^{-1/2})^{\substack{\nu_i\eta_i-(2 \alpha_i' + 2 \beta_i' + 2 \mu_i + 2 \gamma_i) +(\alpha_i'\mu_i - \beta_i' \gamma_i )-(\alpha_i' \mu_i \nu_i +\beta_i' \gamma_i \eta_i)}} \\
& \times (i)^{\substack{\nu_i \eta_i-(\alpha_i' \mu_i \nu_i + \beta_i' \gamma_i \eta_i) +( \alpha_i' \mu_i -\beta_i' \gamma_i)}}\\
& \times (-1)^{\substack{\alpha_i \eta_i +\beta_i \mu_i + \gamma_i \mu_i + \gamma_i \nu_i +\alpha_i' \gamma_i +\alpha_i' \eta_i + \alpha_i' \beta_i' + \beta_i' \mu_i}}d^{3}\mathbf{x} d^{3}\mathbf{x'} d^{3}\mathbf{u}.
\end{aligned}}
\end{equation*}
Summing over the binary components $\alpha_i$, $\beta_i$, $\alpha_i'$, $\beta_i'$, $\mu_i$, $\gamma_i$, $\nu_i$, and $\eta_i$ gives a polynomial in the phase factors $e^{iu}$, $e^{iv}$, $e^{iw}$, $e^{ix}$, $e^{iy}$, $e^{iz}$, $e^{ix'}$, $e^{iy'}$, and $e^{iz'}$, given in Appendix~B. Introduce $e^{iu}=\omega_1$, $e^{iv}=\omega_3$, $e^{iw}=\omega_2$, $e^{ix}=\omega_4$, $e^{iy}=\omega_6$, $e^{iz}=\omega_5$, $e^{ix'}=\omega_7$, $e^{iy'}=\omega_9$, and $e^{iz'}=\omega_8$. We obtain

\begin{equation*}
\mathfit{56}{\begin{aligned}
& \Psi^{m,n,k,m',n',k'}_{p,q,r}=\frac{1}{2^{2N}}\frac{1}{(2\pi)^9}
\Int_{\vert\omega_1\vert=1}\cdots \\
& \times \Int_{\vert\omega_9\vert=1}
\frac{\left(\begin{gathered}
4-\frac{2\omega_1\omega_2\omega_5\omega_6}{3}
-\frac{2\omega_1\omega_2\omega_8\omega_9}{3}\\
+\frac{4\omega_7\omega_9}{3}+\frac{4\omega_2\omega_3}{3}+\cdots
\end{gathered}\right)^N}{\begin{gathered}
\omega_1^{p+1}\omega_2^{q+1}\omega_3^{r+1}\\
\times\omega_4^{m+1}\omega_5^{k+1}\omega_6^{n+1}\\
\times\omega_7^{m'+1}\omega_8^{k'+1}\omega_9^{n'+1}
\end{gathered}} d\omega_1\cdots d\omega_9
\end{aligned}}
\end{equation*}
To reduce the polynomial, make the change of variables

\begin{equation*}
\mathfit{57}{\begin{aligned}
& \omega_1^2=\frac{\Omega_1 \Omega_2}{\Omega_3}, \quad \omega_2^2=\frac{\Omega_1 \Omega_3}{\Omega_2}, \quad \omega_3^2=\frac{\Omega_2 \Omega_3}{\Omega_1}, \\
& \ldots , \quad \omega_7^2=\frac{\Omega_7 \Omega_8}{\Omega_9}, \quad \omega_8^2=\frac{\Omega_7 \Omega_9}{\Omega_8}, \quad \omega_9^2=\frac{\Omega_8 \Omega_9}{\Omega_7},
\end{aligned}}
\end{equation*}
which gives

\begin{equation*}
\mathfit{58}{\begin{aligned}
& \Psi^{m,n,k,m',n',k'}_{p,q,r}=\frac{1}{2^{2N}}\frac{1}{(2\pi)^9}
\Int_{\vert\Omega_1\vert=1}\cdots \\
& \times \Int_{\vert\Omega_9\vert=1}
\frac{\left(\begin{gathered}
4+\Omega_1\Omega_4\Omega_7+\frac{4\Omega_1\Omega_4\Omega_8}{3}\\
+\frac{4\Omega_1\Omega_4\Omega_9}{3}+\frac{4\Omega_1\Omega_5\Omega_8}{3}+\cdots
\end{gathered}\right)^N}{\begin{gathered}
\Omega_1^{\mu_1+1}\Omega_2^{\mu_2+1}\Omega_3^{\mu_3+1}\\
\times\Omega_4^{\mu_4+1}\Omega_5^{\mu_5+1}\Omega_6^{\mu_6+1}\\
\times\Omega_7^{\mu_7+1}\Omega_8^{\mu_8+1}\Omega_9^{\mu_9+1}
\end{gathered}} d\Omega_1\cdots d\Omega_9.
\end{aligned}}
\end{equation*}
where

\[
\begin{aligned}
\mu_1&=(p-q+r)/2,&\mu_2&=(p+q-r)/2,\\
\mu_3&=(-p+q+r)/2,&\mu_4&=(m-n+k)/2,\\
\mu_5&=(m+n-k)/2,&\mu_6&=(-m+n+k)/2,\\
\mu_7&=(m'-n'+k')/2,&\mu_8&=(m'+n'-k')/2,\\
\mu_9&=(-m'+n'+k')/2.
\end{aligned}
\]
The integrals can then be evaluated using Cauchy's integral formula:

\begin{equation*}
\mathfit{59}{\begin{aligned}
& \int_{C} \frac{f(z)}{(z-z_0)^{n+1}} =\frac{2\pi \mathit{i}}{n!} f^{(n)}(z_0)
\end{aligned}}
\end{equation*}
The function to differentiate is the polynomial

\begin{equation*}
\mathfit{60}{\begin{aligned}
& F(\Omega_1, \Omega_2, ... , \Omega_9) = ( 4+ \Omega_1 \Omega_4 \Omega_7 +\frac{4 \Omega_1 \Omega_4 \Omega_8 }{3} \\
& +\frac{4 \Omega_1 \Omega_4 \Omega_9 }{3}+\frac{4 \Omega_1 \Omega_5 \Omega_8}{3}+4 \Omega_2 \Omega_5 \Omega_8  +\cdots )^N.
\end{aligned}}
\end{equation*}
The resulting expression for $\Psi^{m,n,k,m',n',k'}_{p,q,r}$ is

\begin{equation}
\mathfit{61}{\begin{aligned}
& \Psi^{m,n,k,m',n',k'}_{p,q,r} = \frac{1}{2^{2N}}\frac{1}{(2 \pi)^{9} } \\
& \times \left[  \frac{2 \pi i}{\mu_1!} \frac{d^{\mu_1}}{d\Omega_{1}^{\mu_1}}   \cdots \frac{2 \pi i}{\mu_9!} \frac{d^{\mu_9}}{d\Omega_9^{\mu_9}} \right] F(\Omega_1, ... , \Omega_9)\Big|_{\Omega_1=0}...\Big|_{\Omega_9=0}
\end{aligned}}\label{eq:s3_7}
\end{equation}
The first three applications of Cauchy's formula, with respect to $\Omega_1$, $\Omega_2$, and $\Omega_3$, can be evaluated directly:

\begin{equation*}
\mathfit{62}{\begin{aligned}
& \frac{d^{\mu_1}}{d \Omega_1^{\mu_1} } \frac{d^{\mu_2}}{d\Omega_2^{\mu_2}} \frac{d^{\mu_3}}{d\Omega_3^{\mu_3} } [ F(\Omega_1, \Omega_2,\Omega_3, ... , \Omega_9)  ] \Big|_{\Omega_1=0} \Big|_{\Omega_2=0} \Big|_{\Omega_3=0} \\
& =  \frac{N!}{ ( \frac{2N-p-q-r}{2} )!}   \mathbf{P}_1 \mathbf{P}_2 \mathbf{P}_3 \mathbf{P}_4,
\end{aligned}}
\end{equation*}
where

\begin{equation}
\mathfit{63}{\begin{aligned}
& \mathbf{P}_1= ( \frac{2 \mathit{i}}{\sqrt{3}} \Omega_6 - \frac{2 \mathit{i}}{\sqrt{3}}\Omega_9 -\frac{2 \mathit{i}}{\sqrt{3}}\Omega_6\,\Omega_8+\frac{2 \mathit{i}}{\sqrt{3}} \Omega_5\,\Omega_9 \\
& +\frac{2 \mathit{i}}{\sqrt{3}}\Omega_8-\frac{2 \mathit{i}}{\sqrt{3}}\Omega_5-\frac{2}{3}\Omega_9 -\frac{2}{3} \Omega_8 -\frac{2}{3} \Omega_5 \Omega_9 \\
& - \frac{2}{3} \Omega_6 \Omega_8+\frac{4}{3}\Omega_6 \Omega_9+\frac{4}{3} \Omega_4 \Omega_8+4 \Omega_4 \Omega_7 +\frac{4}{3} \Omega_4 \Omega_9\\
& +\frac{4}{3}\Omega_5 \Omega_8-\frac{2}{3}\Omega_5-\frac{2}{3}\Omega_6 +\frac{4}{3}\Omega_4+\frac{4}{3} )^{\frac{p-q+r}{2}}
\end{aligned}}\label{eq:s3_8}
\end{equation}
\begin{equation}
\mathfit{64}{\begin{aligned}
& \mathbf{P}_2 = -\frac{2 \mathit{i}}{\sqrt {3}} \Omega_4 \Omega_9-\frac{2 \mathit{i}}{\sqrt{3}}\Omega_6 +\frac{2 \mathit{i}}{\sqrt{3}}\Omega_6\Omega_7+\frac{2 \mathit{i}}{\sqrt{3}} \Omega_9\\
& +\frac{2 \mathit{i}}{\sqrt{3}}\Omega_4 -\frac{2 \mathit{i}}{\sqrt{3}}\Omega_7-\frac{2}{3} \Omega_9\frac{2}{3}\Omega_7-\frac{2}{3}\Omega_6 \Omega_7 -\frac{2}{3}\Omega_4 \Omega_9\\
& +\frac{4}{3}\Omega_4 \Omega_7+4 \Omega_5 \Omega_8+\frac{4}{3}\Omega_5 \Omega_9 +\frac{4}{3}\Omega_6 \Omega_9\\
& +\frac{4}{3}\Omega_5 \Omega_7-\frac{2}{3}\Omega_6-\frac{2}{3}\Omega_4 +\frac{4}{3}\Omega_5+\frac{4}{3} )^{\frac{p+q-r}{2}}
\end{aligned}}\label{eq:s3_9}
\end{equation}
\begin{equation}
\mathfit{65}{\begin{aligned}
& \mathbf{P}_3 =( \frac{2 \mathit{i}}{\sqrt{3}}\Omega_4\,\Omega_8-\frac{2 \mathit{i}}{\sqrt{3}} \Omega_4  +\frac{2 \mathit{i}}{\sqrt{3}}\Omega_7-\frac{2 \mathit{i}}{\sqrt{3}}\Omega_8\\
&+\frac{2 \mathit{i}}{\sqrt{3}}\Omega_5 -\frac{2 \mathit{i}}{\sqrt{3}}\Omega_5\,\Omega_7-\frac{2}{3}\,\Omega_8-\frac{2}{3}\Omega_7-\frac{2}{3}\Omega_5\Omega_7 \\
& -\frac{2}{3}\Omega_4\,\Omega_8+4\Omega_6\,\Omega_9+\frac{4}{3}\Omega_4\,\Omega_7+\frac{4}{3}\Omega_5\,\Omega_8 \\
& +\frac{4}{3}\Omega_6\,\Omega_7+\frac{4}{3}\Omega_6\,\Omega_8-\frac{2}{3}\Omega_4-\frac{2}{3}\Omega_5 +\frac{4}{3}\Omega_6+\frac{4}{3} ) ^{\frac{-p+q+r}{2}}
\end{aligned}}\label{eq:s3_10}
\end{equation}
\begin{equation}
\mathfit{66}{\begin{aligned}
& \mathbf{P}_4=( 4+\frac{2 \mathit{i}}{\sqrt{3}}\Omega_4\,\Omega_9-\frac{2 \mathit{i}}{\sqrt{3}}\Omega_6\,\Omega_7 -\frac{2 \mathit{i}}{\sqrt{3}}\Omega_4\,\Omega_8 \\
&+\frac{2 \mathit{i}}{\sqrt{3}}\Omega_6\,\Omega_8 -\frac{2 \mathit{i}}{\sqrt{3}}\Omega_5\,\Omega_9+\frac{2 \mathit{i}}{\sqrt{3}}\Omega_5\,\Omega_7 -\frac{2}{3}\Omega_5\,\Omega_9 \\
& -\frac{2}{3}\Omega_6\,\Omega_8-\frac{2}{3}\Omega_4\,\Omega_9-\frac{2}{3}\Omega_6\,\Omega_7-\frac{2}{3}\Omega_4\,\Omega_8 \\
& -\frac{2}{3}\Omega_5\,\Omega_7+\frac{4}{3}\Omega_4\Omega_7+\frac{4}{3}\Omega_5\Omega_8+\frac{4}{3}\Omega_6\Omega_9 \\
& +\frac{4}{3}\Omega_7+\frac{4}{3}\Omega_9+\frac{4}{3}\Omega_8 )^{\frac{2N-p-q-r}{2}}.
\end{aligned}}\label{eq:s3_11}
\end{equation}
The remaining applications of Cauchy's formula can likewise be evaluated and expressed as sums of binomial coefficients. Rather than writing that expanded result, we give the more compact form of $\Psi^{m,n,k,m',n',k'}_{p,q,r}$:

\begin{equation}
\mathfit{67}{\begin{aligned}
& \Psi^{m,n,k,m',n',k'}_{p,q,r} = \frac{1}{2^{2N}} \frac{N!}{ ( \frac{2N-p-q-r}{2} )!}\frac{d^{\mu_4}}{d\Omega_4^{\mu_4}} \frac{d^{\mu_5}}{d\Omega_5^{\mu_5}}  \frac{d^{\mu_6}}{d\Omega_6^{\mu_6}}  \frac{d^{\mu_7}}{d\Omega_7^{\mu_7}} \\
& \times \frac{d^{\mu_8}}{d\Omega_8^{\mu_8}}  \frac{d^{\mu_9}}{d\Omega_9^{\mu_9}} (   \frac{\mathbf{P}_1 \mathbf{P}_2 \mathbf{P}_3 \mathbf{P}_4}{\mu_4!\mu_5!\mu_6!\mu_7!\mu_8!\mu_9!}) \Big|_{\Omega_4=0} \cdots \Big|_{\Omega_9=0}
\end{aligned}}\label{eq:s3_12}
\end{equation}
The kernel $\Psi_{p,q,r}^{m,n,k,m',n',k'}$ is calculated numerically only once for each $N$ and can then be used for any state $\hat{\rho}$ and Hamiltonian $\hat{H}$.

Thus, for two operators, here $\hat{\rho}$ and $\hat{H}$, the projected $Q$ function is calculated as

\begin{equation}
\mathfit{68}{\begin{aligned}
& \tilde{Q}_{\rho H}(p,q,r) = \sum_{\substack{m, n, k, \\ m', n', k'}}  \frac{\tilde{Q}_{\rho}(m,n,k) P_{H}(m',n',k')}{R_{m,n,k}} \Psi_{p,q,r}^{m,n,k,m',n',k'}
\end{aligned}}\label{eq:s3_13}
\end{equation}
When $\hat{H}$ is not permutation-invariant, we propose the approximation

\begin{equation}
\mathfit{69}{\begin{aligned}
& \tilde{Q}_{\rho H}(p,q,r) = \sum_{\substack{m, n, k, \\ m', n', k'}}  \frac{\tilde{Q}_{\rho}(m,n,k) \tilde{P}_{H}(m',n',k')}{R_{m,n,k} R_{m',n',k'}} \Psi_{p,q,r}^{m,n,k,m',n',k'}
\end{aligned}}\label{eq:s3_14}
\end{equation}
Exchanging the two operators replaces $\Psi$ by its complex conjugate $\bar{\Psi}$, giving

\begin{equation}
\mathfit{70}{\begin{aligned}
& \tilde{Q}_{H \rho}(p,q,r) = \sum_{\substack{m, n, k, \\ m', n', k'}}  \frac{\tilde{Q}_{\rho}(m,n,k) \tilde{P}_{H}(m',n',k')}{R_{m,n,k} R_{m',n',k'}} \bar{\Psi}_{p,q,r}^{\, m,n,k,m',n',k'}.
\end{aligned}}\label{eq:s3_15}
\end{equation}
These expressions define an analog of the star product in Eq.~\eqref{eq:s2_33}. Instead of multiplying two $Q$ functions, as in the usual construction, we combine projected $\tilde{Q}$ and $\tilde{P}$ functions:

 \begin{equation}
\mathfit{71}{\begin{aligned}
& \tilde{Q}_{\rho H} = \tilde{Q}_{\rho} \ast \tilde{P}_{H} = \hat{L}( \tilde{Q}_{\rho}, \tilde{P}_{H}),
\end{aligned}}\label{eq:s3_16}
\end{equation}
where

  \begin{equation}
\mathfit{72}{\begin{aligned}
& \hat{L}( \tilde{Q}_{\rho}, \tilde{P}_{H})=\sum_{\substack{m, n, k, \\ m', n', k'}} R_{m,n,k}^{-1} R_{m',n',k'}^{-1}\\
& \times \bar{\Psi}_{p,q,r}^{\, m,n,k,m',n',k'} \tilde{Q}_{\rho}(m,n,k) \tilde{P}_{H}(m',n',k').
\end{aligned}}\label{eq:s3_17}
\end{equation}
Following Sec.~2.7, the evolution equation to solve is, with $\hbar=1$,

\begin{equation}
\mathfit{73}{\begin{aligned}
& \dot{\tilde{Q}}_{\rho}= \mathit{i} [ \tilde{Q}_{\rho H} - \tilde{Q}_{H \rho}]
\end{aligned}}\label{eq:s3_18}
\end{equation}
It can be rewritten as

\begin{equation}
\mathfit{74}{\begin{aligned}
& \dot{\tilde{Q}}_{\rho}(p,q,r)= \mathit{i} \sum_{m,n,k} \Lambda_{p,q,r,m,n,k} \hat{Q}_{\rho}(m,n,k),
\end{aligned}}\label{eq:s3_19}
\end{equation}
where

\begin{equation}
\mathfit{75}{\begin{aligned}
& \Lambda_{p,q,r,m,n,k}= \sum_{m',n',k'}   \frac{ \tilde{P}_{H}(m',n',k')}{R_{m,n,k} R_{m',n',k'}} \\
&\times [\Psi_{p,q,r}^{m,n,k,m',n',k'} - \bar{\Psi}_{p,q,r}^{\,m,n,k,m',n',k'}  ]
\end{aligned}}\label{eq:s3_20}
\end{equation}
To solve Eq.~\eqref{eq:s3_19}, express it in matrix form:

\begin{equation*}
\mathfit{76}{\begin{aligned}
& \frac{d}{dt} \begin{bmatrix}\tilde{Q}_1 \\  \vdots \\ \tilde{Q}_i \\ \vdots \\ \tilde{Q}_M   \end{bmatrix} = \mathit{i}
\begin{bmatrix}
 \Lambda_{1 1} & \Lambda_{1 2} & \ldots & \Lambda_{1 M} \\
\Lambda_{2 1} & \Lambda_{2 2} & \ldots & \Lambda_{2 M}  \\
\vdots & \vdots  & \ddots & \vdots \\
\Lambda_{M 1} &\Lambda_{M 2} & \ldots  & \Lambda_{M M}
\end{bmatrix}
\begin{bmatrix}\tilde{Q}_1 \\ \vdots \\ \tilde{Q}_i \\ \vdots \\ \tilde{Q}_M   \end{bmatrix}
\end{aligned}}
\end{equation*}
The index $i$ specifies an ordering of the triples $(m,n,k)$ through a rule $i\longleftrightarrow(m,n,k)$. The number of points inside the tetrahedron, and hence the vector dimension, is $M=(N+1)(N+2)(N+3)/6$.

The evolution is therefore

\begin{equation*}
\mathfit{77}{\begin{aligned}
& \begin{bmatrix}\tilde{Q}_1(t) \\ \tilde{Q}_2(t) \\ \vdots \\ \tilde{Q}_M(t)   \end{bmatrix} =
 \mathsf{A}
 \begin{bmatrix}
 e^{\mathit{i} t \lambda_{1}} & 0 & \ldots & 0 \\
0& e^{\mathit{i} t \lambda_{2}}  & \ldots & 0  \\
\vdots & \vdots  & \ddots & \vdots \\
0 &0 & \ldots  & e^{\mathit{i} t \lambda_{M}}
\end{bmatrix}
 \quad\times \mathsf{A}^{-1}
\begin{bmatrix}\tilde{Q}_1(t=0) \\ \tilde{Q}_2(t=0) \\ \vdots \\ \tilde{Q}_M(t=0)   \end{bmatrix},
\end{aligned}}
\end{equation*}
Here $\mathbbm{A}$ is the matrix of eigenvectors of the matrix with elements $\Lambda_{i,j}$, and $\lambda_i$ are its eigenvalues.

\section{Results and discussion}
\subsection{Hellinger distance}
The information preserved or lost during evolution with the proposed method can be assessed using the Hellinger distance \cite{Hellinger} between exact and approximate evolution. This measure compares two distribution functions and quantifies their proximity in measurement space.

For two functions $\tilde{Q}_1(m,n,k)$ and $\tilde{Q}_2(m,n,k)$ in the space of collective measurements, the Hellinger distance is defined by

\begin{equation}
\mathfit{78}{\begin{aligned}
& D^{2}(\tilde{Q}_1, \tilde{Q}_2)=1- \frac{1}{2^N} \sum_{m,n,k} \sqrt{\tilde{Q}_1(m,n,k)\tilde{Q}_2(m,n,k)}, \hspace{0.5cm} 0\leq  D \leq 1,
\end{aligned}}\label{eq:s4_1}
\end{equation}
When $D=0$, the functions coincide; when $D=1$, they are completely distinct in this measure.

\subsection{Evolution under different Hamiltonians}
Having defined the projected $Q$ function of a product of operators and the proposed evolution method, we examine the previously introduced states under several permutation-invariant Hamiltonians.

We begin with the Harper Hamiltonian, used to study magnetic properties of atoms in an optical lattice \cite{Harper}. It takes the form

\begin{equation}
\mathfit{79}{\begin{aligned}
& \hat{H}=\hat{J}_x+\hat{J}_y+\hat{J}_z,
\end{aligned}}\label{eq:s4_2}
\end{equation}
The first state is GHZ, Eq.~\eqref{eq:s2_24}. In measurement space, it has regions of higher density at opposite ends of the tetrahedron. Because these two density maxima are far apart, one might expect the approximation to perform less well for some Hamiltonians than for states that are more localized in measurement space.\\

\begin{figure}[H]
\centering
\begin{subfigure}[t]{.89\linewidth}
\centering
\includegraphics[width=\linewidth]{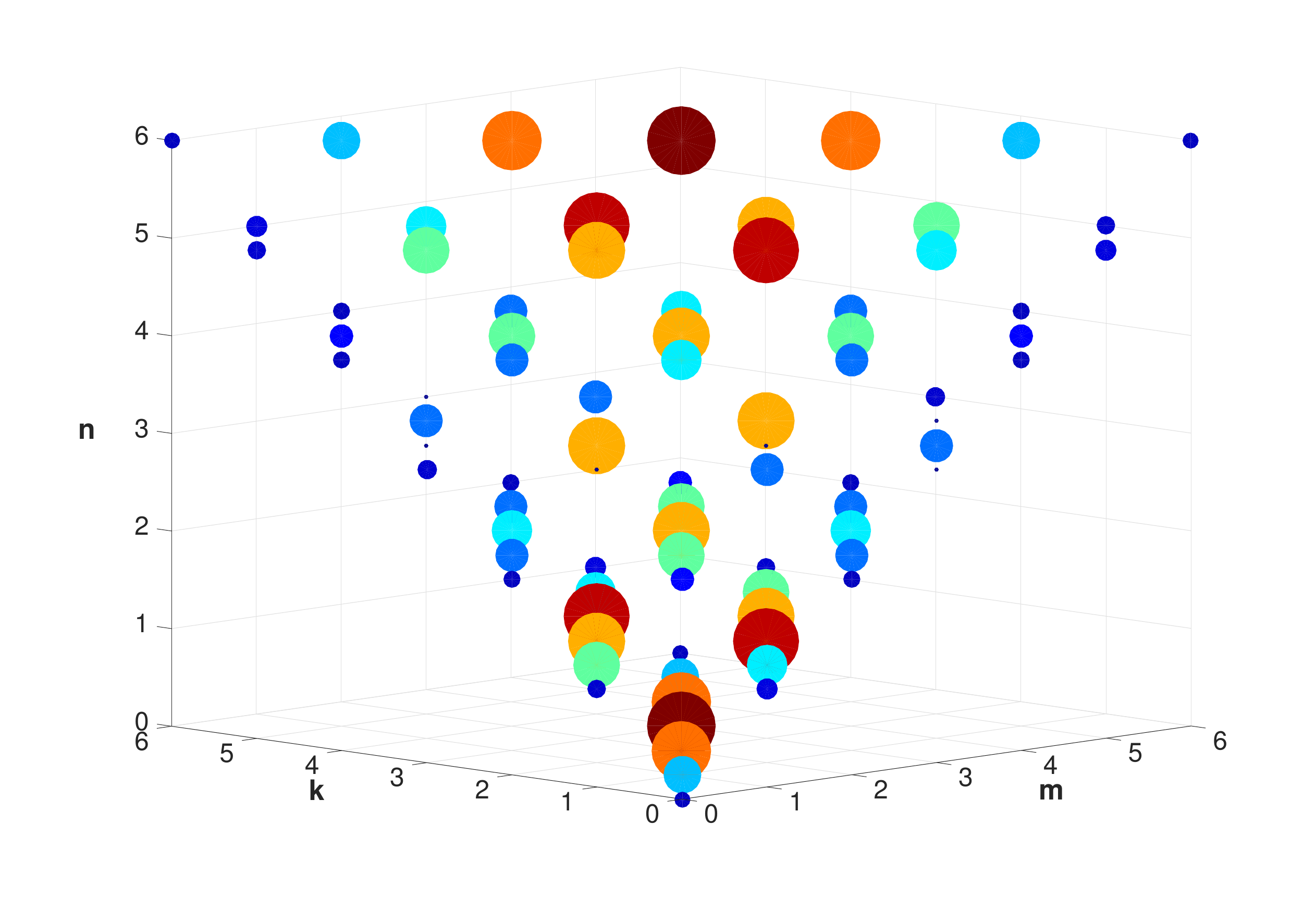}
\caption{}
\end{subfigure}\par\medskip
\begin{subfigure}[t]{.89\linewidth}
\centering
\includegraphics[width=\linewidth]{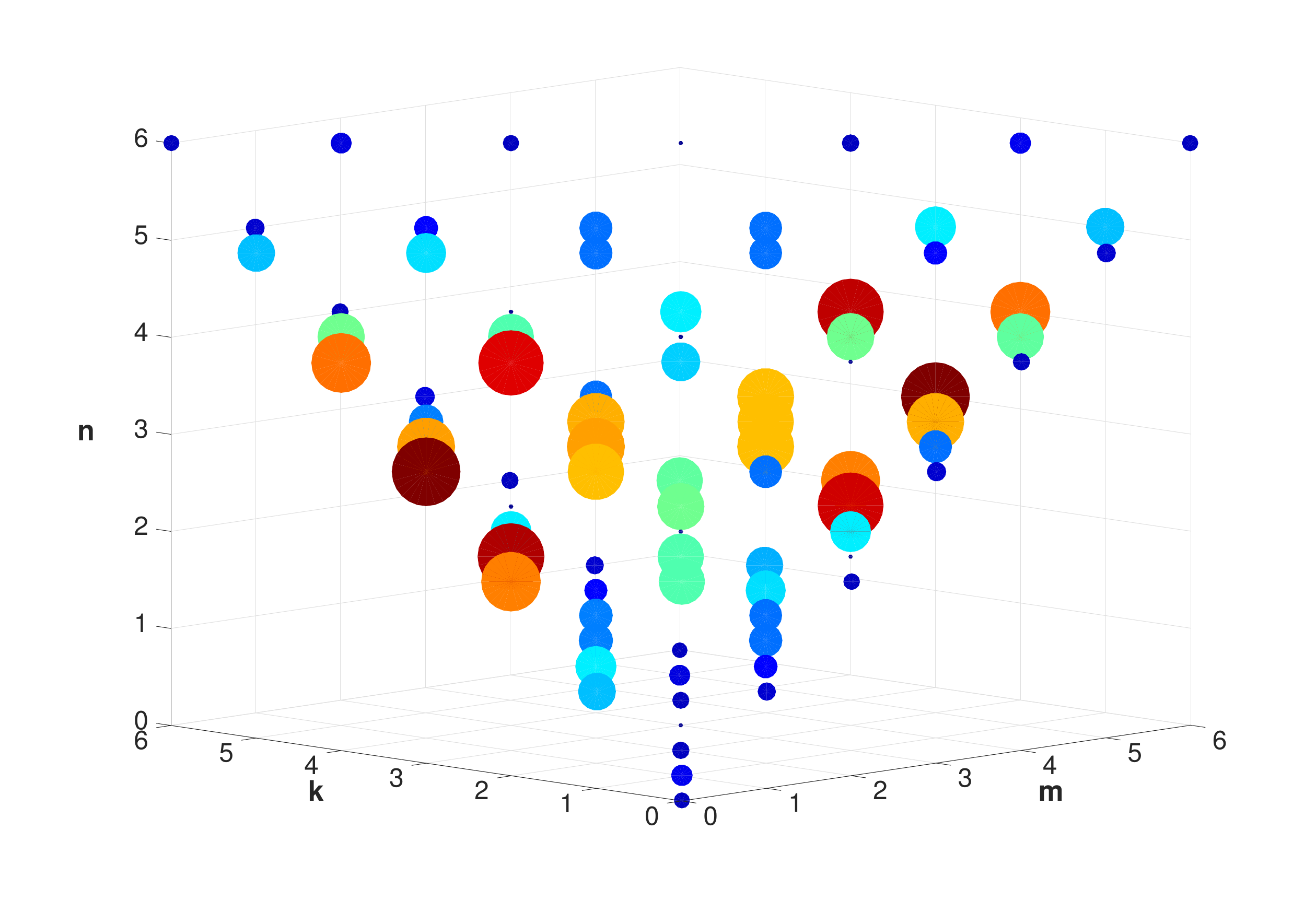}
\caption{}
\end{subfigure}\par\medskip
\begin{subfigure}[t]{.89\linewidth}
\centering
\includegraphics[width=\linewidth]{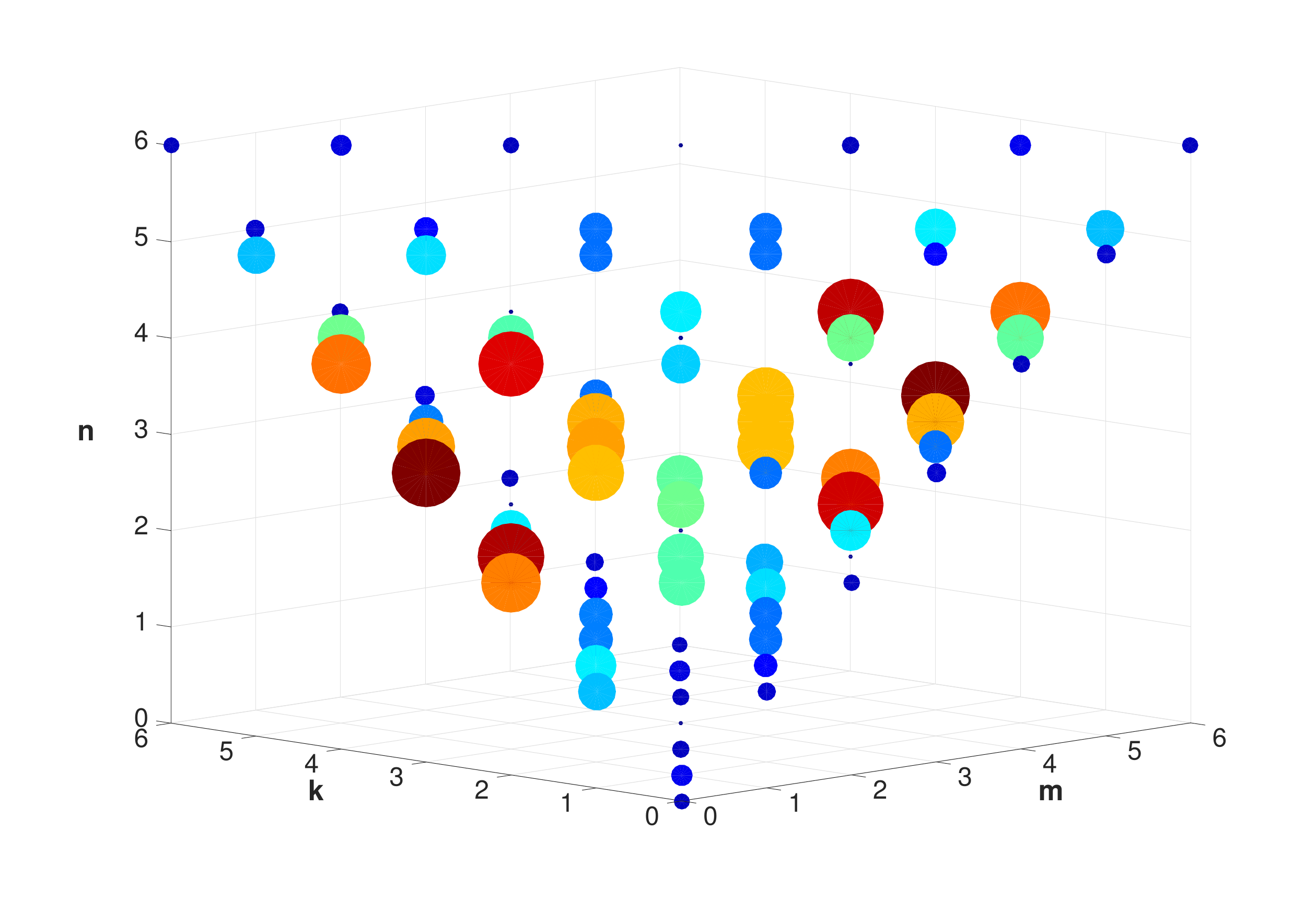}
\caption{}
\end{subfigure}
\caption{GHZ projected distributions: (a) initial $\tilde{Q}_{GHZ}$ at $t=0$; (b) approximate evolution at $t=0.6$; and (c) exact evolution at $t=0.6$, under the Hamiltonian in Eq.~\eqref{eq:s4_2}.}
\label{fig:s4_1}
\end{figure}

Fig.~\ref{fig:s4_1}(a) shows the projected GHZ $Q$ function at $t=0$ in the tetrahedral measurement space described in Sec.~2.4.1.

Fig.~\ref{fig:s4_1}(b) and 4.1(c) show the distributions obtained under the Hamiltonian in Eq.~\eqref{eq:s4_2}. The approximate function is practically identical to the exact-evolution result. Thus, the proposed approximation performs very well for this case, where both the state and Hamiltonian are permutation-invariant.\\

During evolution, the initially separated concentrations move along the $n$ axis, merge, and then separate along the $m$--$k$ direction. Ultimately, the points exhibit mirror symmetry with respect to the $m$ and $k$ axes.

Another case of interest is a Kerr medium, which can generate squeezing of coherent states at short times \cite{Kerr}.

For the Kerr medium, take the Hamiltonian

\begin{equation}
\mathfit{80}{\begin{aligned}
& \hat{H}=\hat{J}_z^2,
\end{aligned}}\label{eq:s4_3}
\end{equation}
and consider the measurement-space evolution of the coherent state in Eq.~\eqref{eq:s2_26}.\\

\begin{figure}[H]
\centering
\begin{subfigure}[t]{.89\linewidth}
\centering
\includegraphics[width=\linewidth]{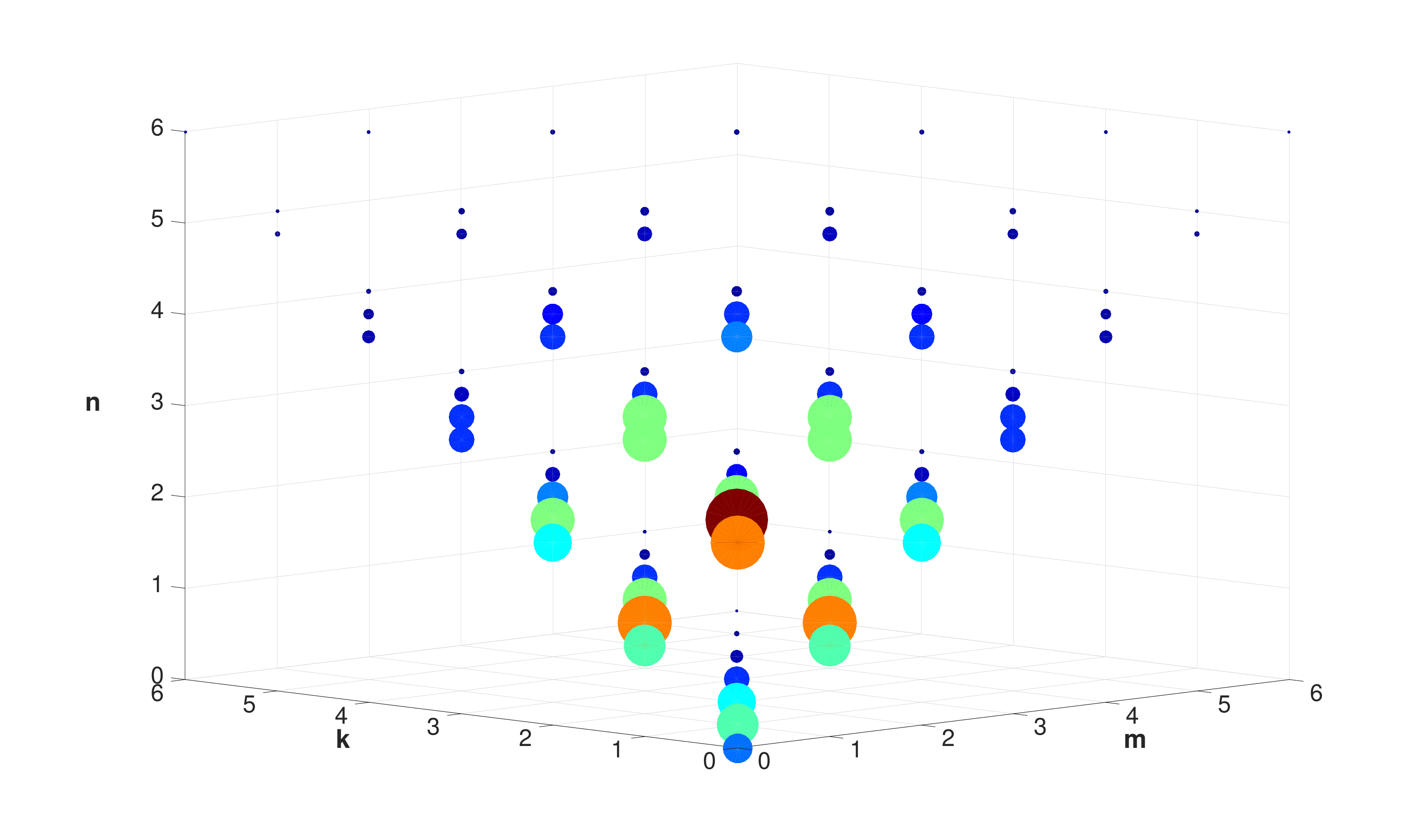}
\caption{}
\end{subfigure}\par\medskip
\begin{subfigure}[t]{.89\linewidth}
\centering
\includegraphics[width=\linewidth]{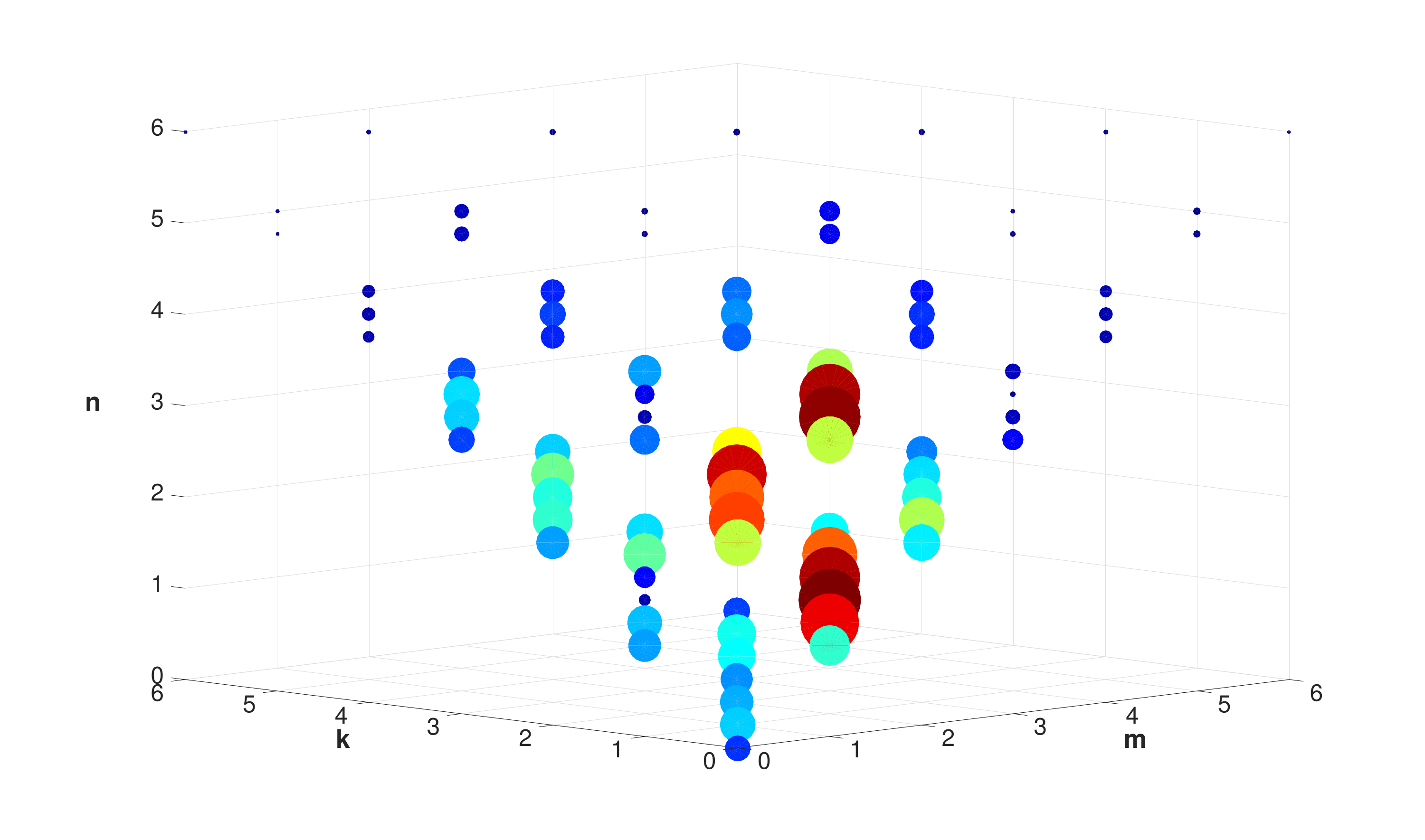}
\caption{}
\end{subfigure}\par\medskip
\begin{subfigure}[t]{.89\linewidth}
\centering
\includegraphics[width=\linewidth]{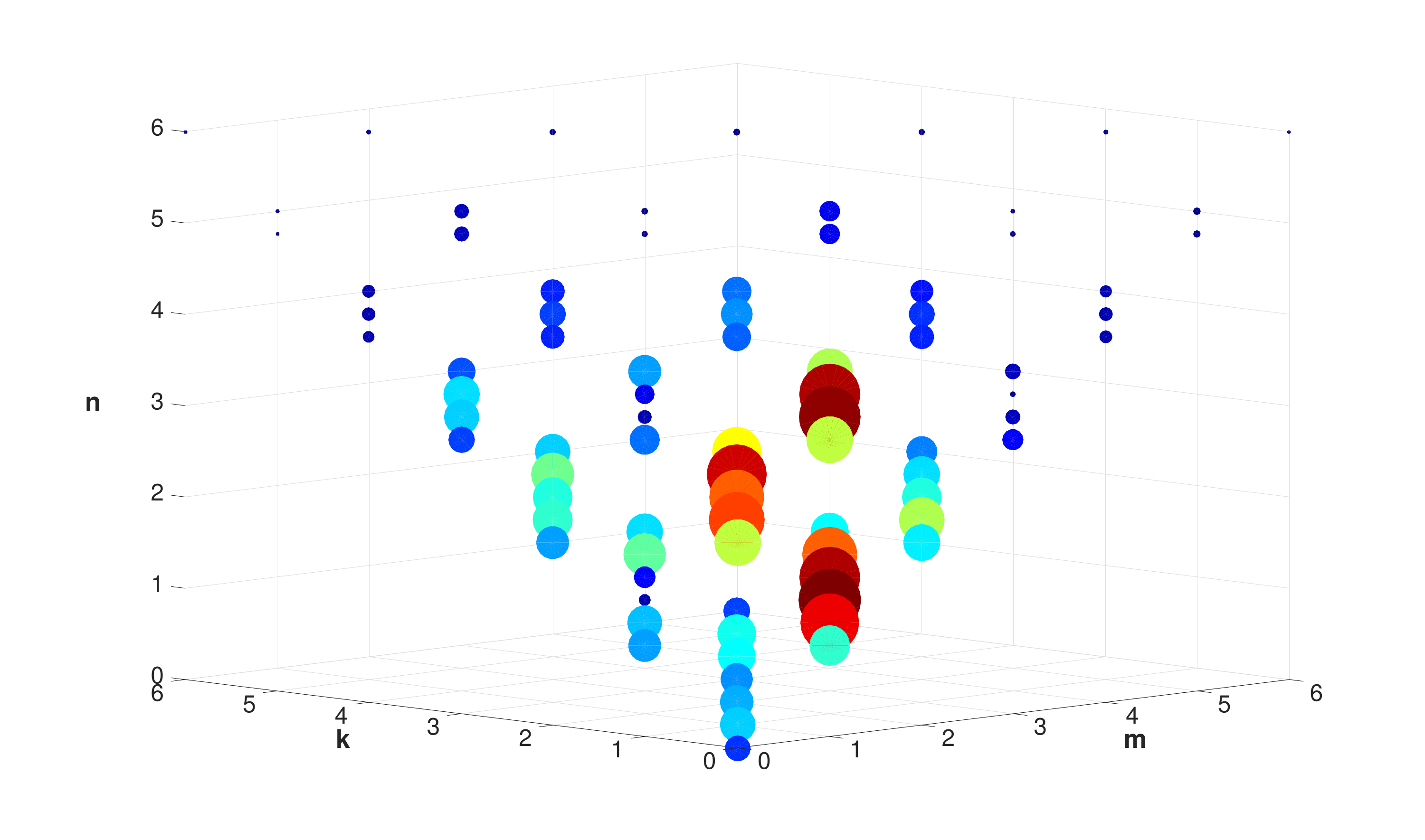}
\caption{}
\end{subfigure}
\caption{Coherent-state projected distributions: (a) initial $\tilde{Q}_{\xi}$ at $t=0$; (b) approximate evolution at $t=0.4$; and (c) exact evolution at $t=0.4$, under the Hamiltonian in Eq.~\eqref{eq:s4_3}.}
\label{fig:s4_2}
\end{figure}

Fig.~\ref{fig:s4_2}, like Fig.~\ref{fig:s4_1}, compares approximate and exact evolution of the projected $Q$ function, here under Eq.~\eqref{eq:s4_3}. Again the two functions are very similar, confirming good performance for a permutation-invariant Hamiltonian and state.

The information concentrations move toward the $n$ axis, including a concentration containing more points of high density than at $t=0$.

After the permutation-invariant examples in Figs.~4.1 and 4.2, we turn to a noninvariant state and the Lipkin--Meshkov--Glick (LMG) model, which arises in nuclear physics \cite{LMG}.

The LMG Hamiltonian is

\begin{equation}
\mathfit{81}{\begin{aligned}
& \hat{H}=-\hat{J}_z^2-g \hat{J}_x.
\end{aligned}}\label{eq:s4_4}
\end{equation}
Consider the domain-wall state, Eq.~\eqref{eq:s2_27}, which is not permutation-invariant, evolving under Eq.~\eqref{eq:s4_4}.\\

\begin{figure}[H]
\centering
\begin{subfigure}[t]{.89\linewidth}
\centering
\includegraphics[width=\linewidth]{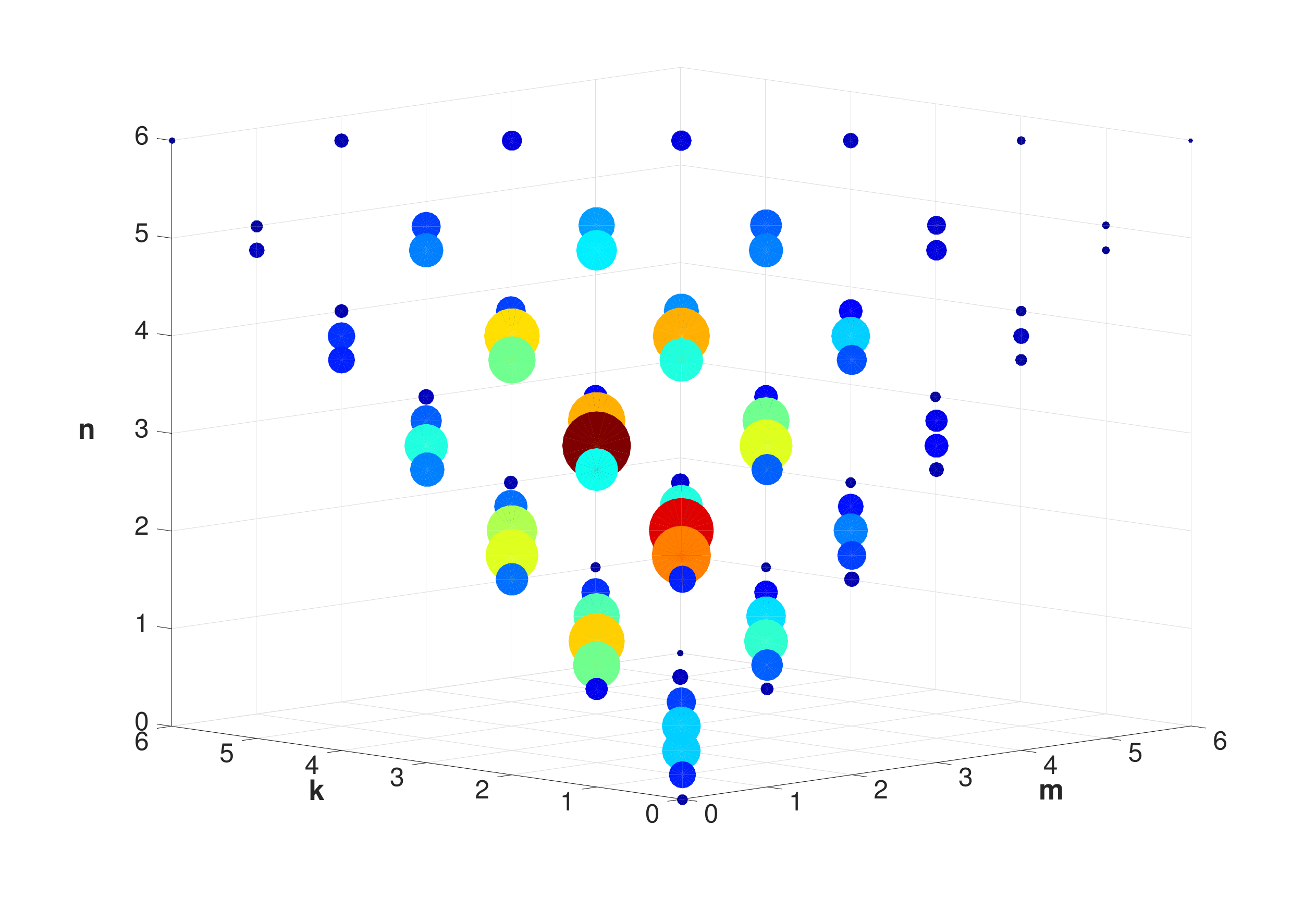}
\caption{}
\end{subfigure}\par\medskip
\begin{subfigure}[t]{.89\linewidth}
\centering
\includegraphics[width=\linewidth]{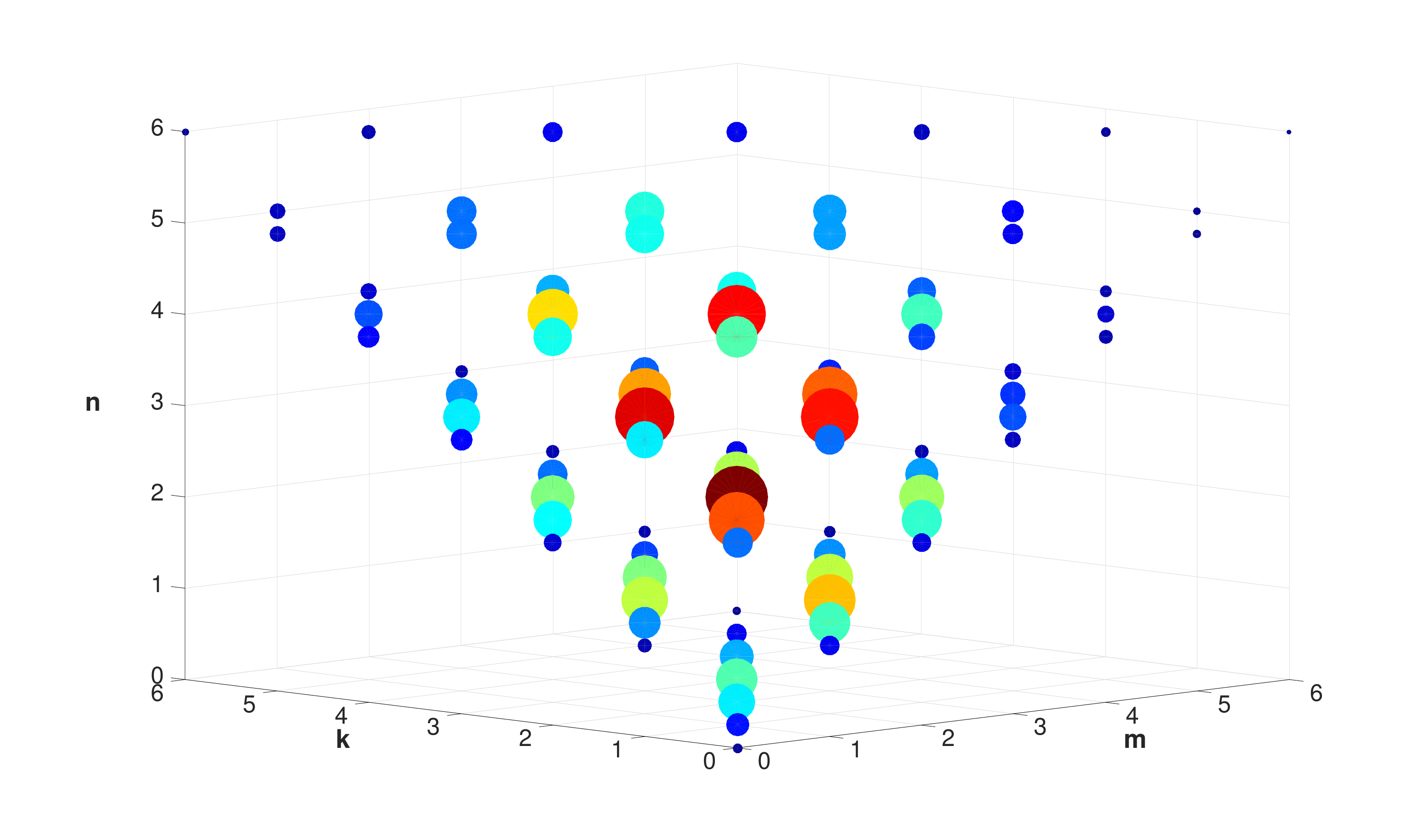}
\caption{}
\end{subfigure}\par\medskip
\begin{subfigure}[t]{.89\linewidth}
\centering
\includegraphics[width=\linewidth]{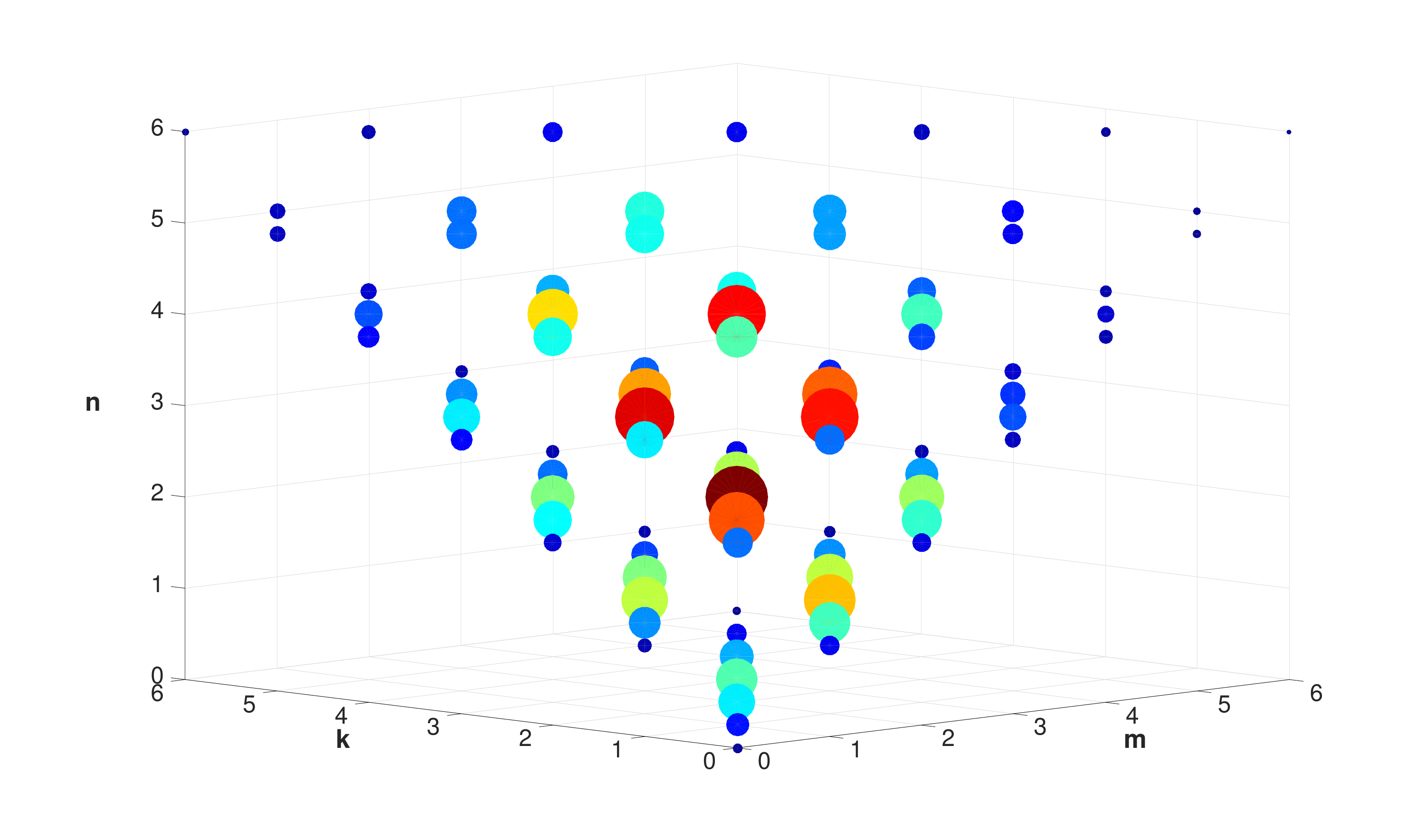}
\caption{}
\end{subfigure}
\caption{Domain-wall projected distributions for $g=1$: (a) initial $\tilde{Q}_{DW}$ at $t=0$; (b) approximate evolution at $t=0.5$; and (c) exact evolution at $t=0.5$, under Eq.~\eqref{eq:s4_4}.}
\label{fig:s4_3}
\end{figure}

Fig.~\ref{fig:s4_3}(b) and 4.3(c) again compare approximate and exact projected distributions. Interestingly, they are very similar after evolution, as in the permutation-invariant examples. Although the initial distribution is spread throughout measurement space, a small region of higher information density is visible near the tetrahedron's center, slightly displaced to the left along $k$. Under Eq.~\eqref{eq:s4_4}, this evolves into a better-defined concentration at the tetrahedron's center.

Thus, under a symmetric Hamiltonian, the approximation works well not only for some permutation-invariant states but also for some noninvariant states.

The previously defined Hellinger distance provides another way to test the approximation for both types of states evolving under a permutation-invariant Hamiltonian.

\begin{figure}[H]
\centering
\begin{subfigure}[t]{1\linewidth}
\centering
\includegraphics[width=\linewidth]{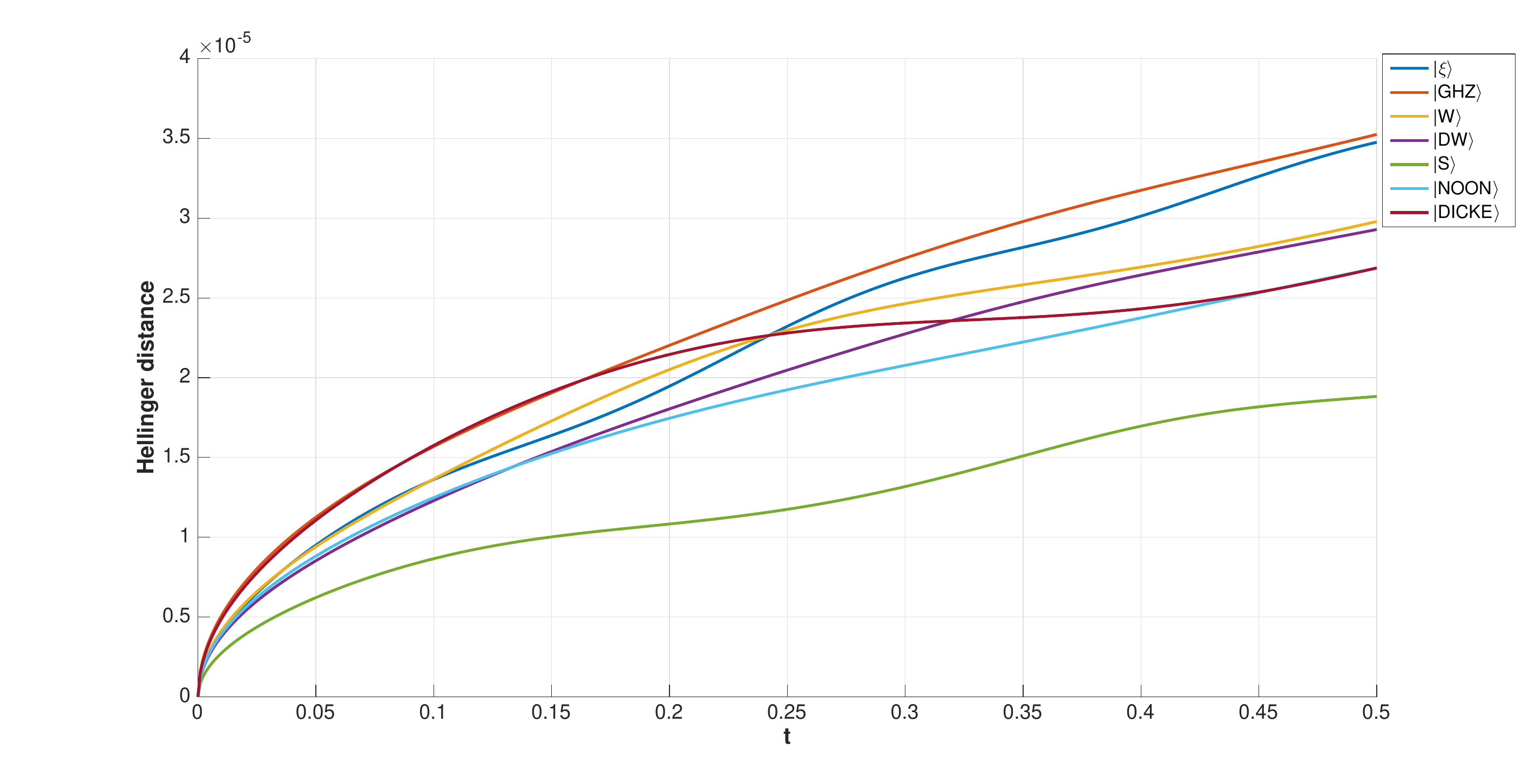}
\end{subfigure}
\caption{Hellinger distance under the LMG Hamiltonian in Eq.~\eqref{eq:s4_4}.}
\label{fig:s4_4}
\end{figure}

As a first example, Fig.~\ref{fig:s4_4} shows the Hellinger distances for several permutation-invariant and noninvariant states under Eq.~\eqref{eq:s4_4}.\\

In Fig.~\ref{fig:s4_4}, the Hellinger distances for both types of states are of order $10^{-5}$, indicating a very small error. Similar results were obtained for the Harper Hamiltonian, Eq.~\eqref{eq:s4_2}, the Kerr Hamiltonian, Eq.~\eqref{eq:s4_3}, and

\begin{equation}
\mathfit{82}{\begin{aligned}
& \hat{H}=\hat{J}_x^2+\hat{J}_y^2+\hat{J}_z^2,
\end{aligned}}\label{eq:s4_5}
\end{equation}
The corresponding distance plots are therefore not shown; for some states the distances are of order $10^{-10}$.

As discussed in the methodology, the method was expected to perform well for permutation-invariant states and Hamiltonians. However, Fig.~\ref{fig:s4_3} also shows very small errors for non-invariant states. This result indicates that the method remains accurate for non-invariant states when the Hamiltonian is permutation-invariant. Consequently, the approximation

\begin{equation*}
\mathfit{83}{\begin{aligned}
& Q(\alpha,\beta) \approx \tilde{Q}(m,n,k) R_{m,n,k}^{-1},
\end{aligned}}
\end{equation*}
is useful for studying the evolution of quantum states.\\

Although the method is not expected to perform equally well for Hamiltonians that are not permutation-invariant, we now examine several such examples.

We begin with the Ising Hamiltonian in an external field \cite{Ising},

\begin{equation}
\mathfit{84}{\begin{aligned}
& \hat{H}=-[ \sum_i^N \sigma_z^{(i)} \sigma_z^{(i+1)}-\hat{J}_x ].
\end{aligned}}\label{eq:s4_6}
\end{equation}
For the GHZ state evolving under Eq.~\eqref{eq:s4_6}, the projected $Q$ function is shown in Figs.~4.5 and 4.6.\\

Compared with Figs.~4.5 and 4.6, the approximate distribution remains close to the exact distribution in a relatively short time $t=0.3$. In the longer time $t=1.5$, information has been lost at certain points relative to the exact evolution.\\

Now consider a second Ising-type Hamiltonian,

\begin{equation}
\mathfit{85}{\begin{aligned}
& \hat{H}=- [\sum_i^N \sigma_z^{(i)} \sigma_z^{(i+1)} + \sum_j^N \sigma_y^{(j)} \sigma_y^{(j+1)}-\hat{J}_x].
\end{aligned}}\label{eq:s4_7}
\end{equation}
and again examine the GHZ state, now under Eq.~\eqref{eq:s4_7}.\\

Figure~\ref{fig:s4_7} shows dynamics similar to those in Fig.~\ref{fig:s4_1}. In both cases, concentrations initially located near the ends of the $n$ axis move toward the center and subsequently occupy symmetric positions along the $m$ and $k$ axes. Although the Hamiltonian in Eq.~\eqref{eq:s4_7} is not permutation invariant, the approximation remains accurate at this relatively long evolution time: the concentrations occupy similar positions and have comparable values, as indicated by the sizes and colors of the spheres.

Nevertheless, the approximation does not accurately capture all dynamics generated by the Hamiltonian in Eq.~\eqref{eq:s4_7}. As an example, we next consider the evolution of a coherent state.\\

Fig.~\ref{fig:s4_11} supports a similar analysis: some noninvariant states appear to perform as well as invariant states, which was unexpected. For each state, the Hellinger distances grow similarly under the two Hamiltonians. In particular, the singlet-state distances appear to take almost the same values in both cases.

A further result from the LMG model concerns its quantum phase transition.

A phase transition is an abrupt change in a system's macroscopic characteristics as one of its defining variables changes. The points at which these changes occur are critical points and generally mark the transition from one type of behavior, or phase, to another \cite{Valtierra}.

Fig.~\ref{fig:s4_8} and 4.9 display behavior similar to Figs.~4.5 and 4.6. At short times, approximate and exact evolution produce very similar projected distributions. In the longer time in Fig.~\ref{fig:s4_9}, the distributions of the projected $Q$ values in the measurement space are completely different.

\begin{figure}[H]
\centering
\begin{subfigure}[t]{.89\linewidth}
\centering
\includegraphics[width=\linewidth]{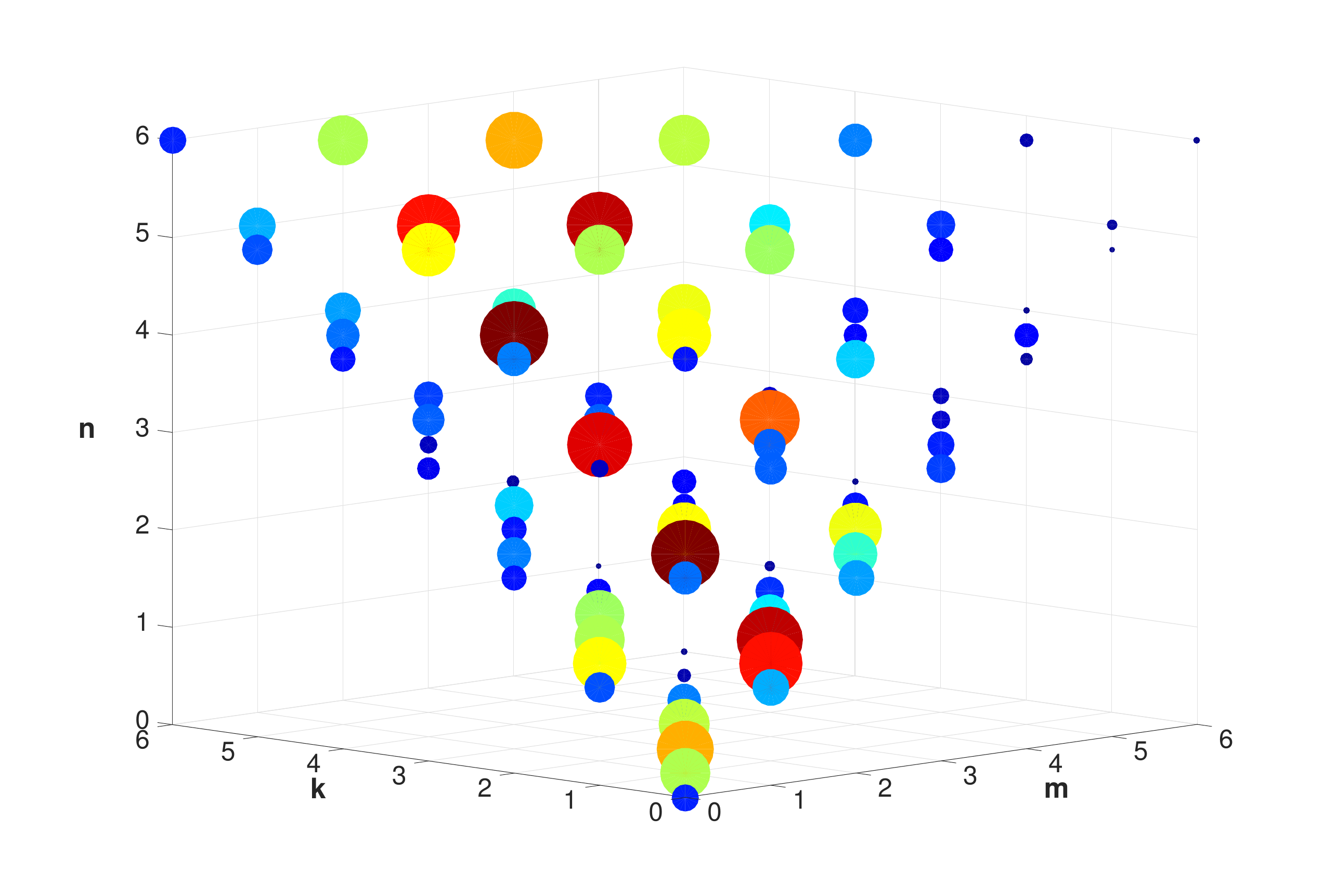}
\caption{}
\end{subfigure}\par\medskip
\begin{subfigure}[t]{.89\linewidth}
\centering
\includegraphics[width=\linewidth]{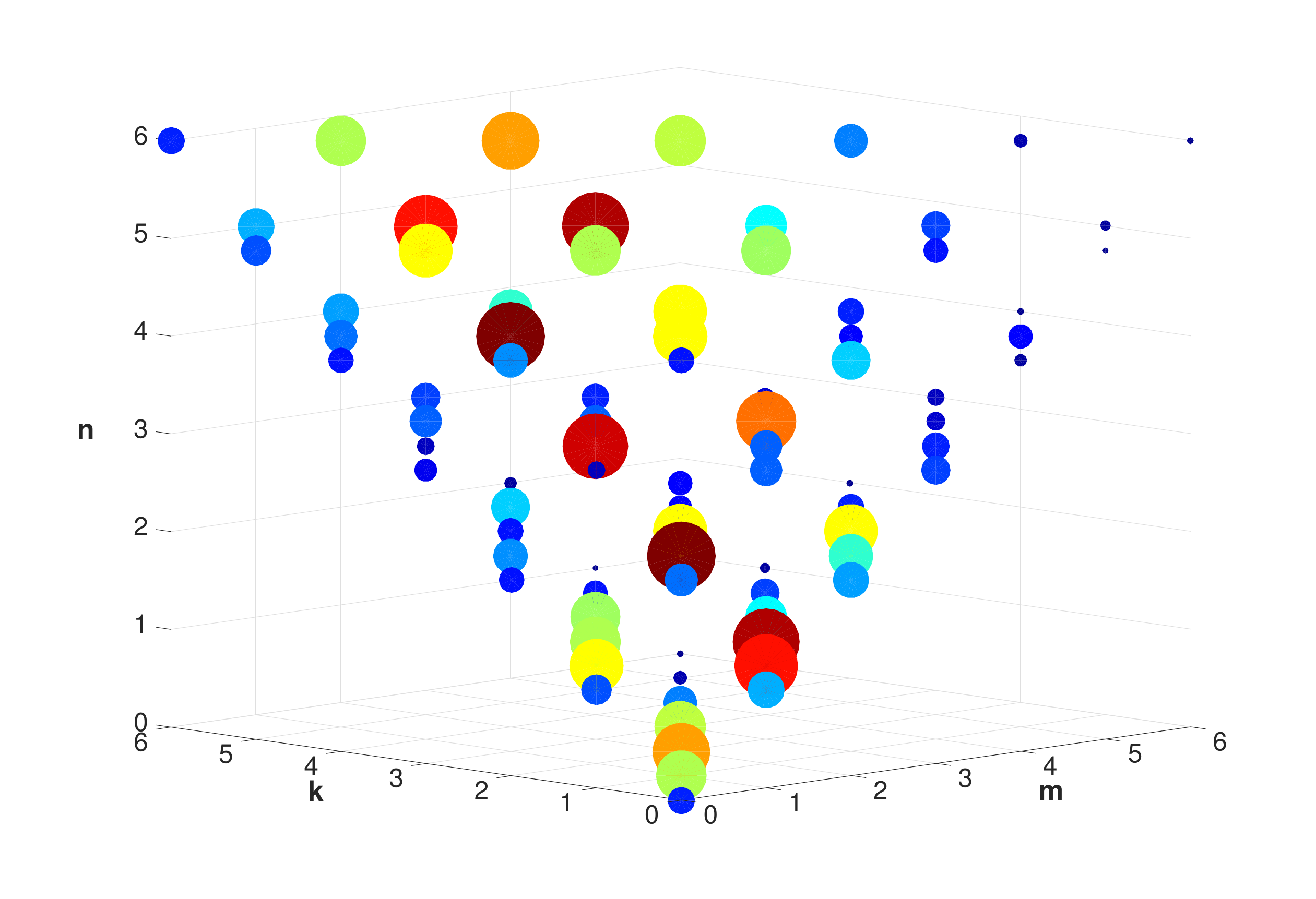}
\caption{}
\end{subfigure}
\caption{GHZ projected distributions under Eq.~\eqref{eq:s4_6} at $t=0.3$: (a) approximate evolution and (b) exact evolution.}
\label{fig:s4_5}
\end{figure}

\begin{figure}[H]
\centering
\begin{subfigure}[t]{.89\linewidth}
\centering
\includegraphics[width=\linewidth]{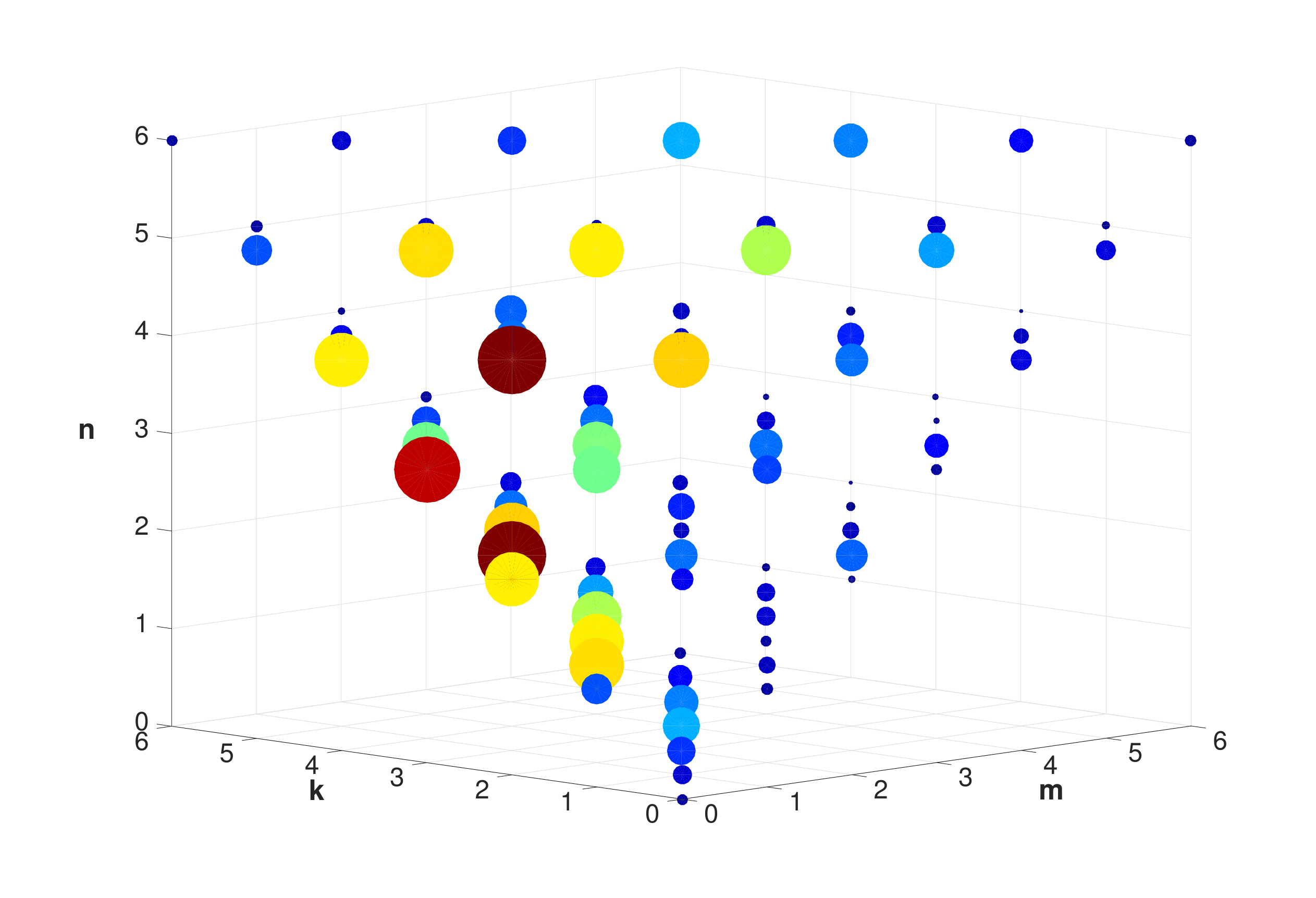}
\caption{}
\end{subfigure}\par\medskip
\begin{subfigure}[t]{.89\linewidth}
\centering
\includegraphics[width=\linewidth]{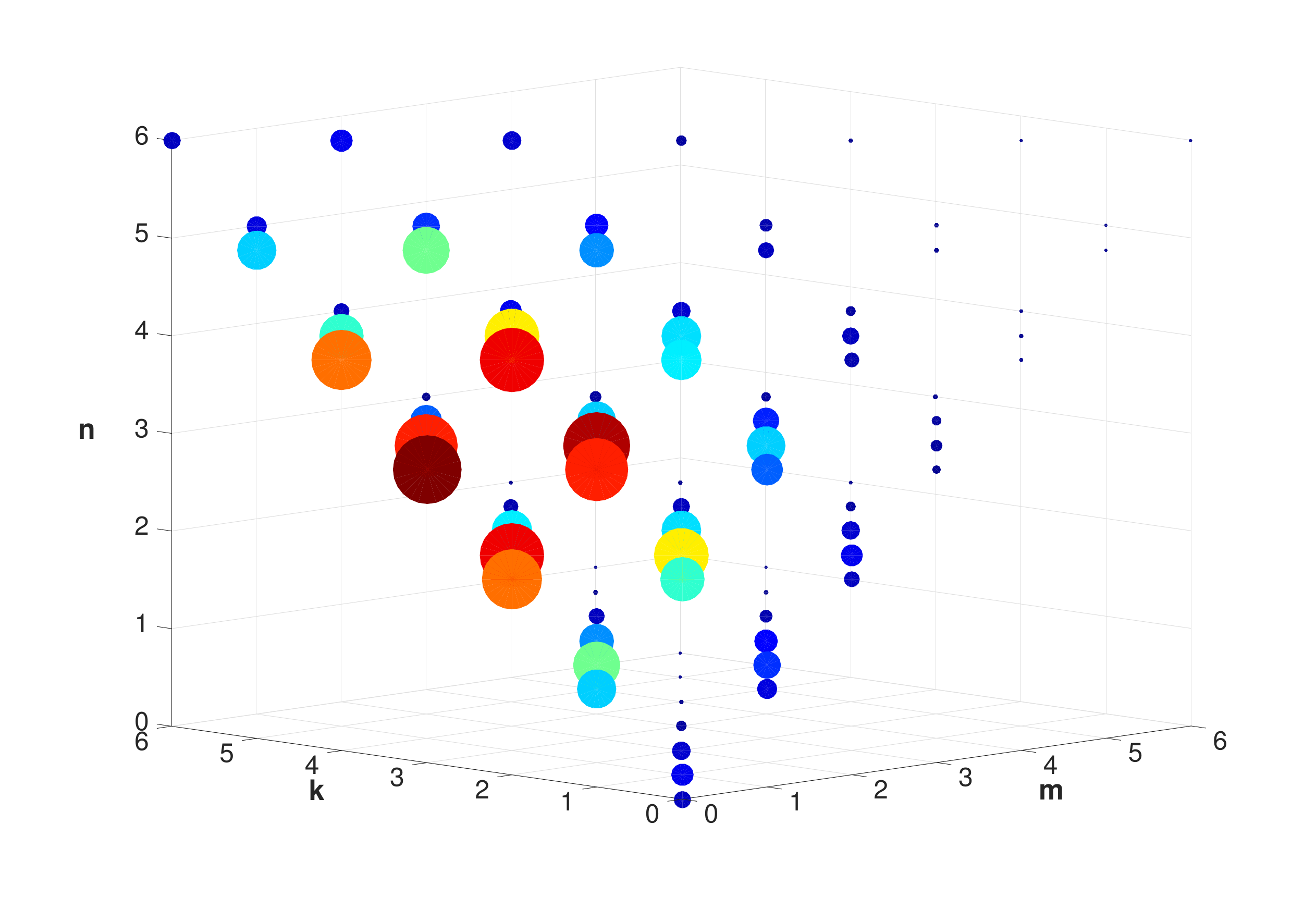}
\caption{}
\end{subfigure}
\caption{GHZ projected distributions under Eq.~\eqref{eq:s4_6} at $t=1.5$: (a) approximate evolution and (b) exact evolution.}
\label{fig:s4_6}
\end{figure}

\begin{figure}[H]
\centering
\begin{subfigure}[t]{.89\linewidth}
\centering
\includegraphics[width=\linewidth]{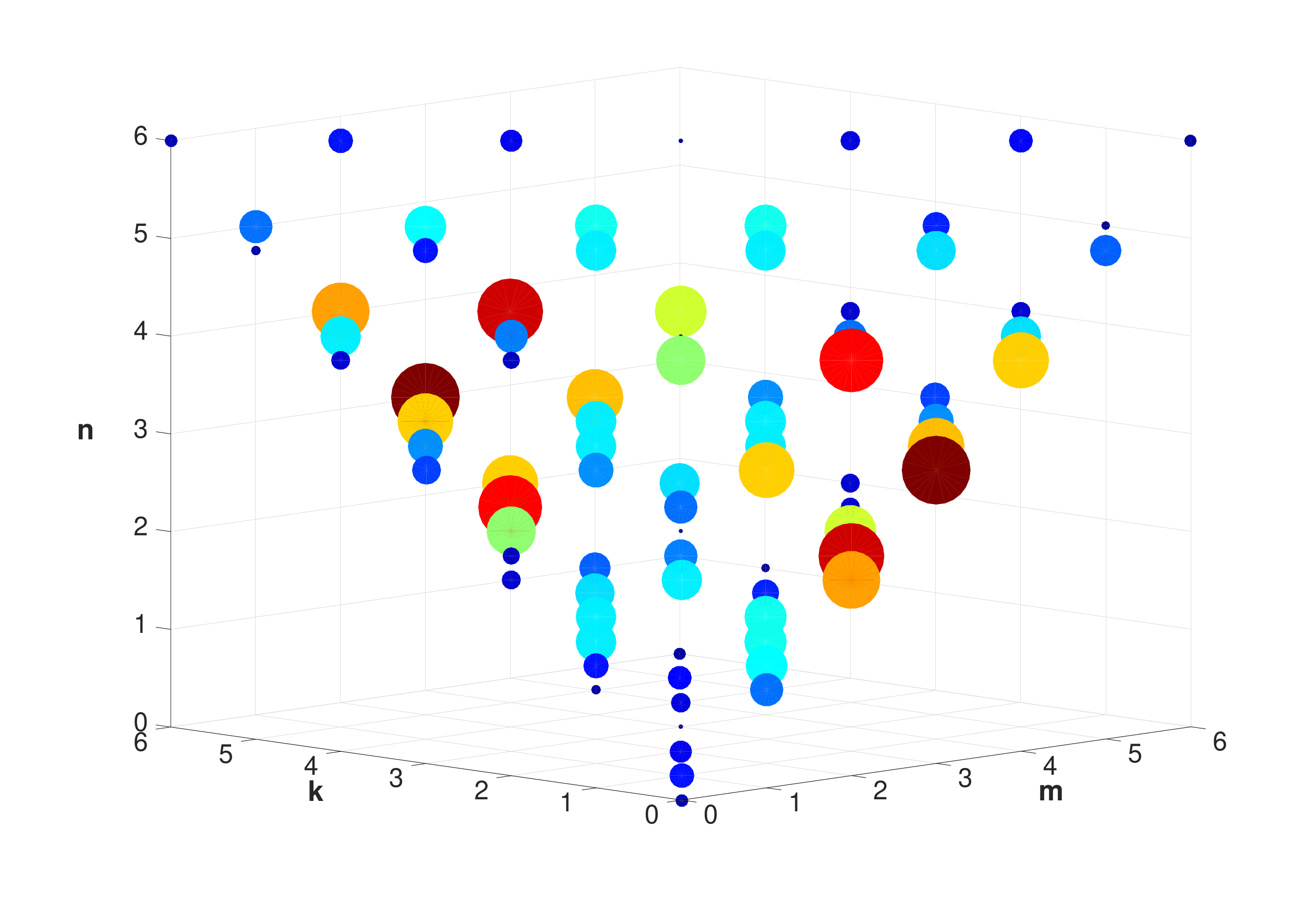}
\caption{}
\end{subfigure}\par\medskip
\begin{subfigure}[t]{.89\linewidth}
\centering
\includegraphics[width=\linewidth]{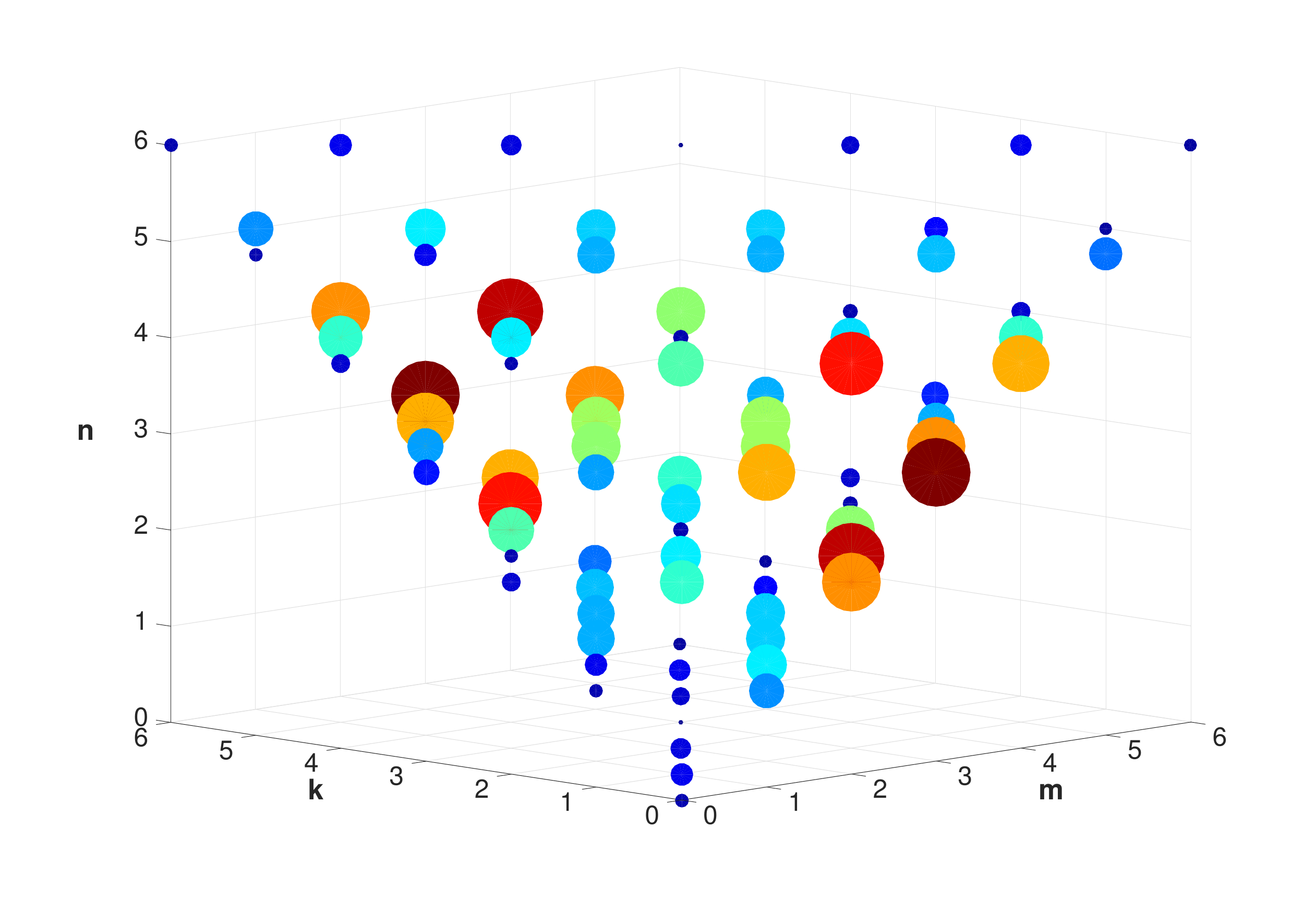}
\caption{}
\end{subfigure}
\caption{GHZ projected distributions under Eq.~\eqref{eq:s4_7} at $t=0.8$: (a) approximate evolution and (b) exact evolution.}
\label{fig:s4_7}
\end{figure}

\begin{figure}[H]
\centering
\begin{subfigure}[t]{.89\linewidth}
\centering
\includegraphics[width=\linewidth]{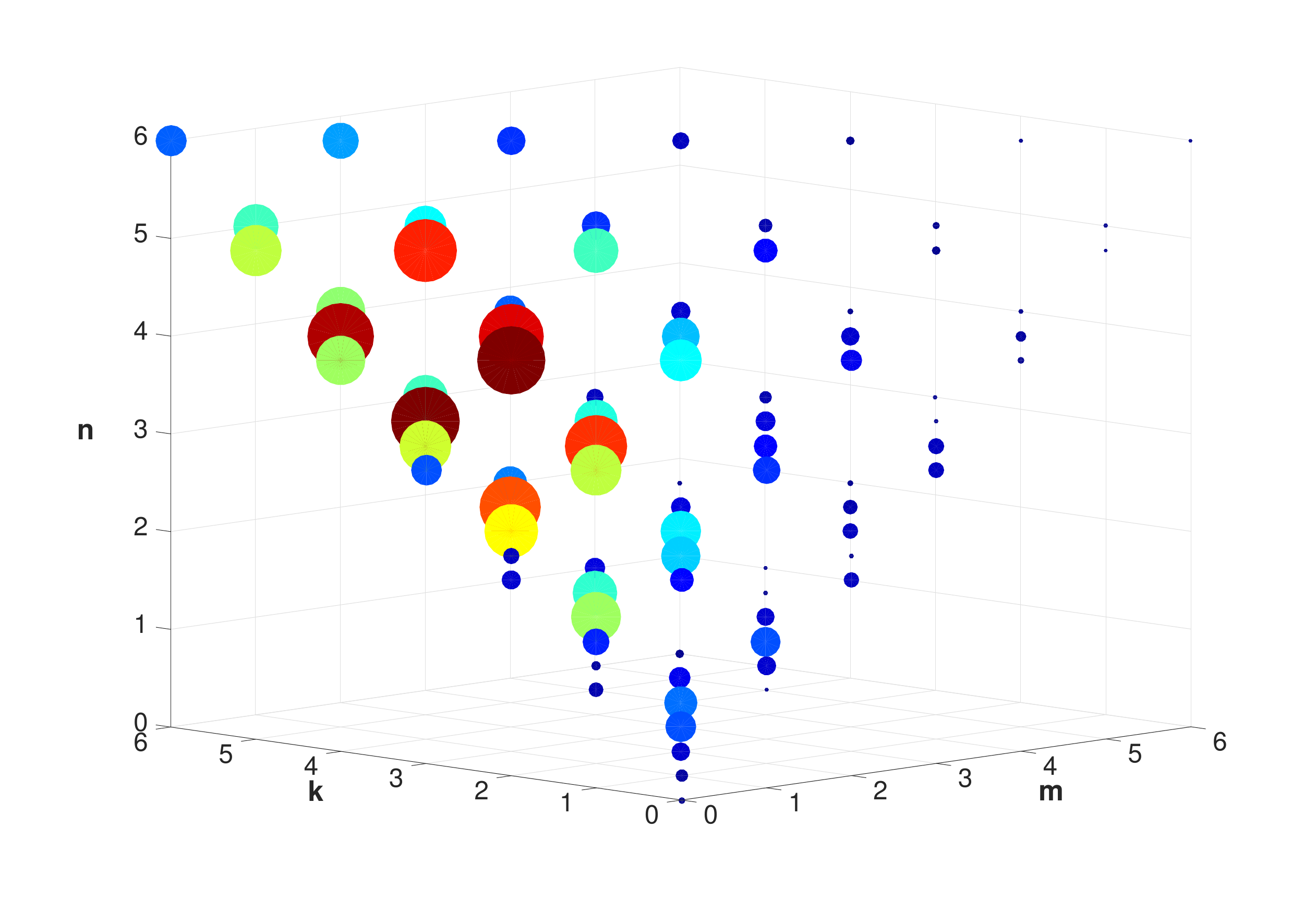}
\caption{}
\end{subfigure}\par\medskip
\begin{subfigure}[t]{.89\linewidth}
\centering
\includegraphics[width=\linewidth]{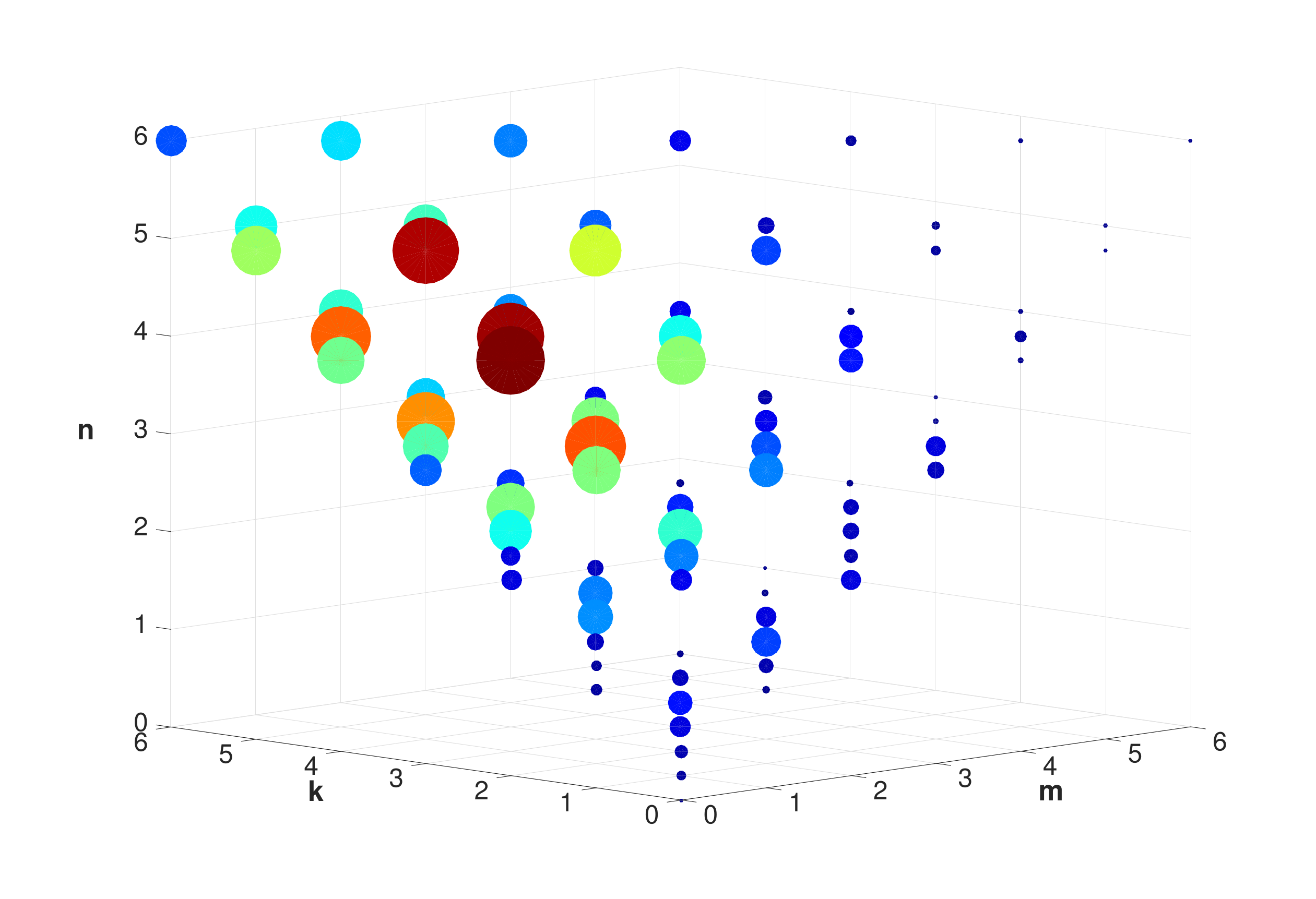}
\caption{}
\end{subfigure}
\caption{Coherent-state projected distributions under Eq.~\eqref{eq:s4_7} at $t=0.6$: (a) approximate evolution and (b) exact evolution.}
\label{fig:s4_8}
\end{figure}
\begin{figure}[H]
\centering
\begin{subfigure}[t]{.89\linewidth}
\centering
\includegraphics[width=\linewidth]{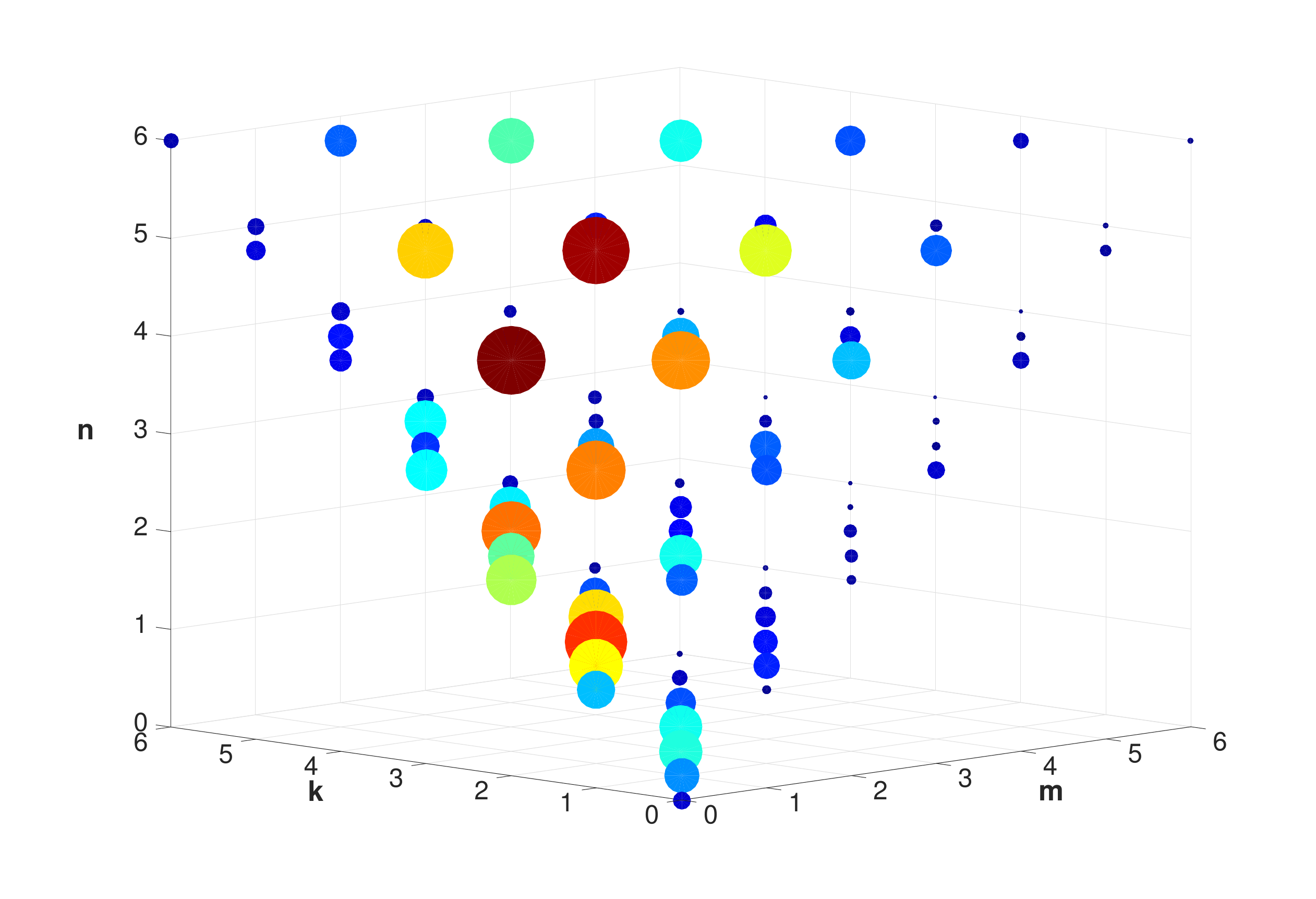}
\caption{}
\end{subfigure}\par\medskip
\begin{subfigure}[t]{.89\linewidth}
\centering
\includegraphics[width=\linewidth]{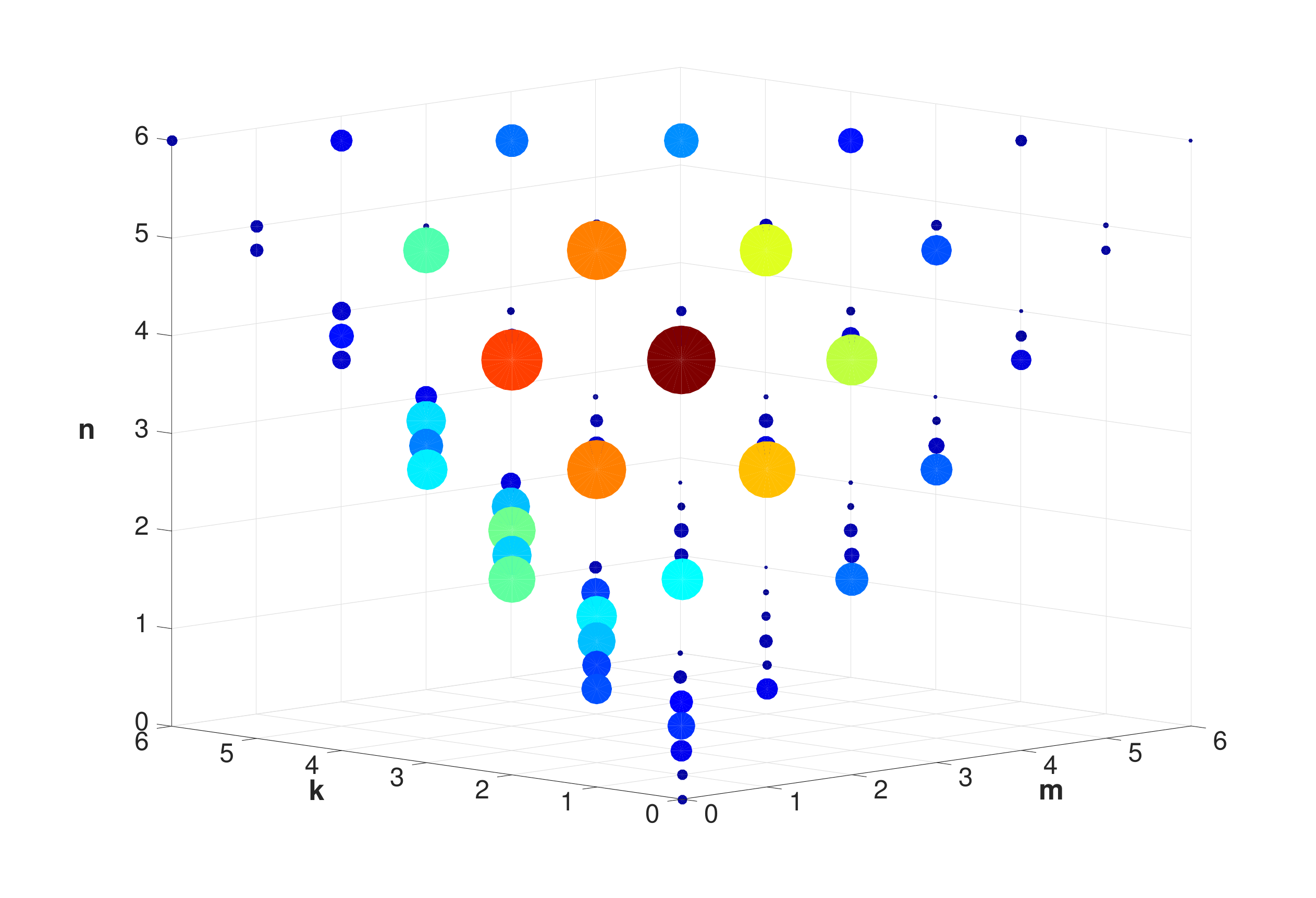}
\caption{}
\end{subfigure}
\caption{Coherent-state projected distributions under Eq.~\eqref{eq:s4_7} at $t=2$: (a) approximate evolution and (b) exact evolution.}
\label{fig:s4_9}
\end{figure}

As for the permutation-invariant Hamiltonians, the Hellinger distances for Eqs.~\eqref{eq:s4_6} and \eqref{eq:s4_7} can be calculated for several initial states.

\begin{figure}[H]
\centering
\begin{subfigure}[t]{1\linewidth}
\centering
\includegraphics[width=\linewidth]{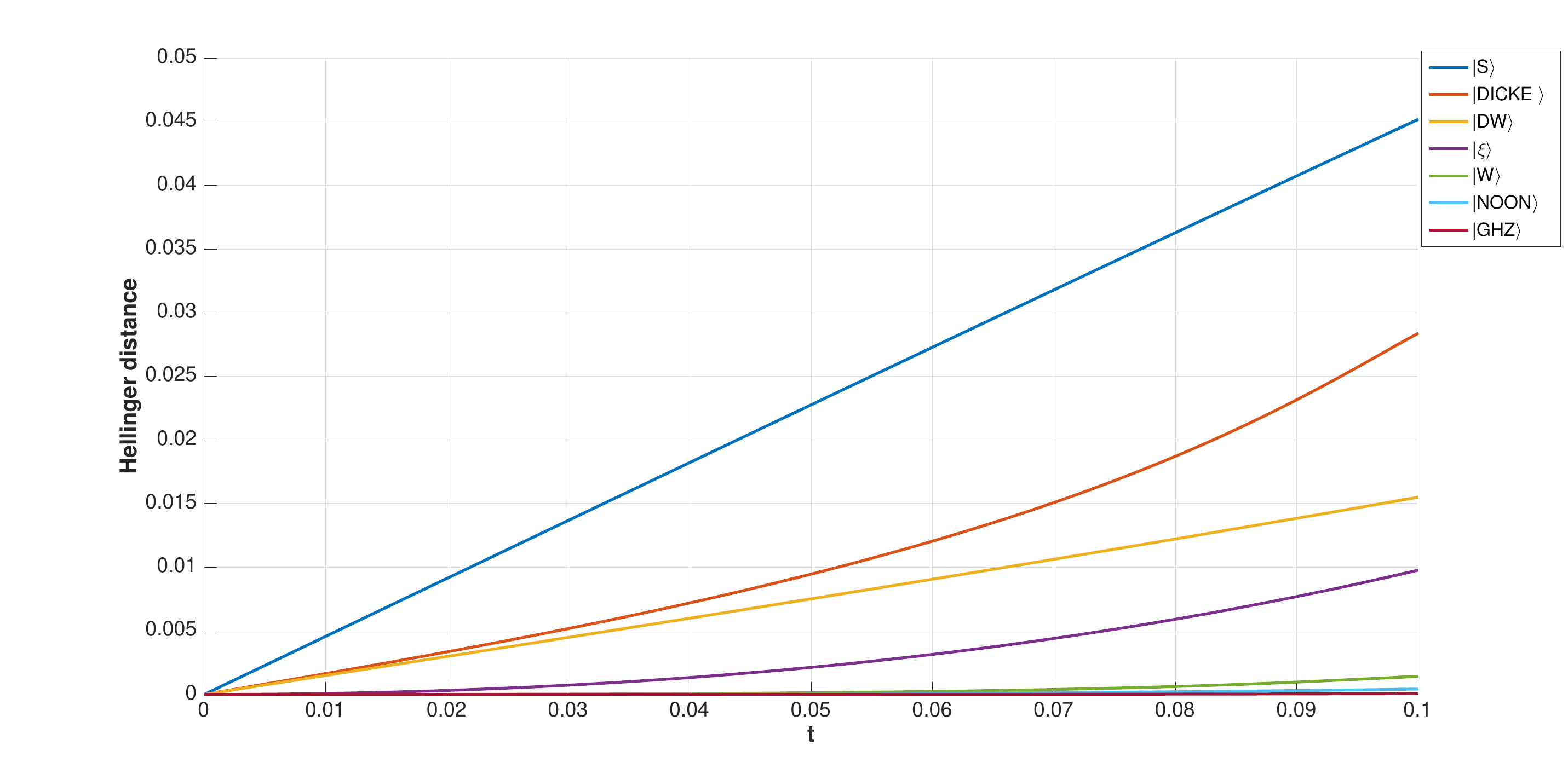}
\end{subfigure}
\caption{Hellinger distance under the Ising Hamiltonian in Eq.~\eqref{eq:s4_6}.}
\label{fig:s4_10}
\end{figure}
Fig.~\ref{fig:s4_10} shows that, under Eq.~\eqref{eq:s4_6}, the distance reaches values larger than those in Fig.~\ref{fig:s4_4} even at very short times. Although better results are generally expected for permutation-invariant states, the NOON state performs better than the coherent state of Eq.~\eqref{eq:s2_26}.

\begin{figure}[H]
\centering
\begin{subfigure}[t]{1\linewidth}
\centering
\includegraphics[width=\linewidth]{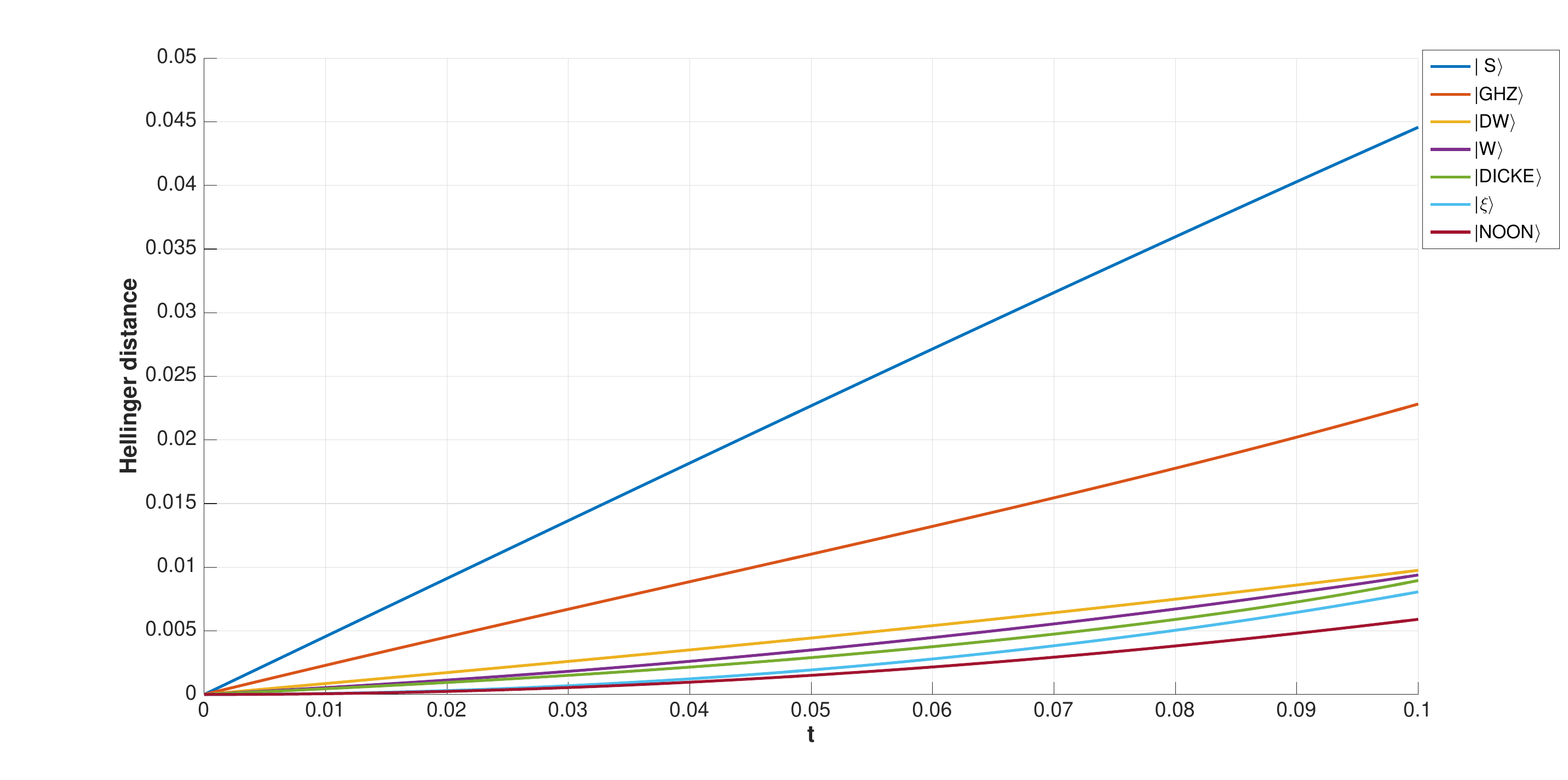}
\end{subfigure}
\caption{Hellinger distance under the second Ising-type Hamiltonian in Eq.~\eqref{eq:s4_7}.}
\label{fig:s4_11}
\end{figure}

For the LMG model, the transition occurs at $\tilde{g}=g/(2N)=1$. We tested the approximation by evolving the coherent state $\vert\xi=1\rangle$, Eq.~\eqref{eq:s2_7}, under Eq.~\eqref{eq:s4_4} for $\tilde{g}=0.5$, $1$, and $1.5$. The Hellinger distances were similar to those for the other permutation-invariant Hamiltonians: all three cases yielded distances of order $10^{-5}$.

\section{Conclusions}
We obtained an approximate expression for the projected $Q$ function of a product of two operators in terms of their projected $\tilde{Q}$ and $\tilde{P}$ functions, by analogy with the star product between $Q$ functions. The resulting expression was used to solve the Liouville equation numerically in the space of collective measurements and to analyze the dynamics of different states and Hamiltonians.

For permutation-invariant states and Hamiltonians, the approximation produced an acceptable error of order $10^{-5}$ or smaller. The error is measured by the Hellinger distance between the projected distribution obtained from the exact state evolution and the distribution obtained from the approximate evolution. Unexpectedly, noninvariant states also evolved accurately under the approximate star product when the Hamiltonian was permutation-invariant.

For Hamiltonians that are not permutation-invariant, the approximation error is not small for every state at long times. Depending on the Hamiltonian, the approximation may still be useful for some states. The relationship between the initial state and a noninvariant Hamiltonian that permits accurate approximation remains to be investigated. Nevertheless, the approximation performs well at short times for both invariant and noninvariant states.

For the LMG Hamiltonian, whose phase transition is controlled by $g$, one might expect different behavior across the transition at $\tilde{g}=1$. The approximation nevertheless continued to perform well and yielded very small errors over long evolution times.

The method therefore performs well even for the reported $N=6$ calculations. Calculations at smaller $N$ increase the error, suggesting that the approximation may improve for larger particle numbers, $N\gg1$. It worked over long times for both invariant and noninvariant states when the Hamiltonian was permutation-invariant. At short times, it also worked for most of the states studied without requiring Hamiltonian symmetry.

\subsection{Future work}
The main direction for future work is to simplify the kernel $\Psi^{m,n,k,m',n',k'}_{p,q,r}$ in Eq.~\eqref{eq:s3_6}. Although it needs to be computed only once for each particle number, its construction requires substantial computation time, which the present work reports as increasing exponentially with $N$. A simpler expression would make it possible to study larger particle numbers and more clearly reveal the characteristics of the projected $Q$ function.

It would also be interesting to extend the method to prime-power dimensions $p^n$ with $p\ne2$.

\section*{Acknowledgments}
The author thanks the National Council of Science and Technology (CONACYT) and the University of Guadalajara for funding and support for this project. 

\FloatBarrier

\appendix
\makeatletter
\renewcommand{\@seccntformat}[1]{\appendixname~\csname the#1\endcsname:\quad}
\makeatother
\section{The kernel \texorpdfstring{$\Psi^{m,n,k,m\prime,n\prime,k\prime}_{p,q,r}$}{Psi}}
The full derivation used to obtain a more compact expression for $\Psi^{m,n,k,m',n',k'}_{p,q,r}$ is presented below.

Start from the exact expression

\begin{equation}
\mathfit{86}{\begin{aligned}
& \Psi^{m,n,k,m',n',k'}_{p,q,r} =  \sum_{\substack{\alpha,\beta \\ \alpha, \beta \\ \mu, \nu}} \langle \mu, \nu \vert  \Delta_P(\alpha,\beta)\Delta_Q(\alpha',\beta') \mu, \nu \rangle \\
&\times  \delta_{h(\alpha),m} \delta_{h(\beta),n}\delta_{h(\alpha+\beta),k} \delta_{h(\alpha'),m'}\delta_{h(\beta'),n'}\\
& \times \delta_{h(\alpha'+\beta'),k'} \delta_{h(\mu),p} \delta_{h(\gamma),q}\delta_{h(\mu+\gamma),r}
\end{aligned}}\label{eq:A_1}
\end{equation}
The kernel can be expanded as

\begin{equation*}
\mathfit{87}{\begin{aligned}
& \Psi^{m,n,k,m',n',k'}_{p,q,r} =\sum_{\alpha,\beta,\alpha',\beta',\mu,\gamma}\langle \mu, \gamma \vert \Delta_{P}(\alpha,\beta) \vert \alpha', \beta'\rangle \langle \alpha', \beta' \vert  \mu, \gamma \rangle
\end{aligned}}
\end{equation*}
\begin{equation*}
\mathfit{88}{\begin{aligned}
& \times \delta_{h(\alpha),m}\delta_{h(\beta),n}\delta_{h(\alpha+\beta),k} \delta_{h(\alpha'),m'} \\
& \delta_{h(\beta'),n'}\delta_{h(\alpha'+\beta'),k'} \delta_{h(\mu),p}\delta_{h(\gamma),q}\delta_{h(\mu+\gamma),r}.
\end{aligned}}
\end{equation*}
Using the commutation relation in Eq.~\eqref{eq:s2_3} and the state definition in Eq.~\eqref{eq:s2_4}, we rewrite $\vert\alpha',\beta'\rangle$ and apply the operators:

\begin{equation*}
\mathfit{89}{\begin{aligned}
& =\sum_{\alpha,\beta,\alpha',\beta',\mu,\gamma}\langle \mu, \gamma \vert \Delta_{P}(\alpha,\beta) \vert \alpha', \beta'\rangle \langle \xi \vert X_{\beta'} Z_{\alpha'} Z_{\mu} X_{\gamma}\vert \xi \rangle
\end{aligned}}
\end{equation*}
\begin{equation*}
\mathfit{90}{\begin{aligned}
& \times\delta_{h(\alpha),m}\delta_{h(\beta),n}\delta_{h(\alpha+\beta),k} \delta_{h(\alpha'),m'}\delta_{h(\beta'),n'} \\
& \times \delta_{h(\alpha'+\beta'),k'} \delta_{h(\mu),p}\delta_{h(\gamma),q}\delta_{h(\mu+\gamma),r},
\end{aligned}}
\end{equation*}
\begin{equation*}
\mathfit{91}{\begin{aligned}
& =\sum_{\alpha,\beta,\alpha',\beta',\mu,\gamma}\langle \mu, \gamma \vert \Delta_{P}(\alpha,\beta) \vert \alpha', \beta'\rangle \langle \xi \vert Z_{\alpha'+\mu}  X_{\gamma+\beta'}\vert \xi \rangle \chi(\alpha'\beta' +\mu\beta')
\end{aligned}}
\end{equation*}
\begin{equation*}
\mathfit{92}{\begin{aligned}
& \times \delta_{h(\alpha),m}\delta_{h(\beta),n}\delta_{h(\alpha+\beta),k} \delta_{h(\alpha'),m'}\delta_{h(\beta'),n'} \\
& \times \delta_{h(\alpha'+\beta'),k'} \delta_{h(\mu),p}\delta_{h(\gamma),q}\delta_{h(\mu+\gamma),r},
\end{aligned}}
\end{equation*}
Using the relation

\begin{equation}
\mathfit{93}{\begin{aligned}
& \langle \xi \vert X_{\beta} Z_{\alpha} \vert \xi \rangle = (3^{-1/4})^{h(\alpha)+h(\beta)+h(\alpha + \beta)} \\
& \times (\mathit{i}^{-1/2})^{h(\alpha)+h(\beta)-h(\alpha + \beta)}
\end{aligned}}\label{eq:A_2}
\end{equation}
the preceding equation becomes

\begin{equation*}
\mathfit{94}{\begin{aligned}
& =\sum_{\alpha,\beta,\alpha',\beta',\mu,\gamma}\langle \mu, \gamma \vert \Delta_{P}(\alpha,\beta) \vert \alpha', \beta'\rangle \\
& \times (3^{-1/4})^{h(\alpha'+\mu)+h(\beta'+\gamma)+h(\alpha'+\mu+\beta'+\gamma)} \\
& \times (\mathit{i}^{1/2})^{h(\alpha'+\mu)+h(\beta'+\gamma)-h(\alpha'+\mu+\beta'+\gamma)}
\end{aligned}}
\end{equation*}
\begin{equation*}
\mathfit{95}{\begin{aligned}
& \times \chi(\alpha'\beta'+\beta'\mu) \delta_{h(\alpha),m}\delta_{h(\beta),n}\delta_{h(\alpha+\beta),k} \\
& \times \delta_{h(\alpha'),m'}\delta_{h(\beta'),n'}\delta_{h(\alpha'+\beta'),k'}\delta_{h(\mu),p}\delta_{h(\gamma),q} \delta_{h(\mu+\gamma),r},
\end{aligned}}
\end{equation*}
Furthermore, using Eq.~\eqref{eq:s2_11}, the $P$ kernel can be expanded as

\begin{equation*}
\mathfit{96}{\begin{aligned}
& =\sum_{\alpha,\beta,\alpha',\beta',\mu,\gamma}\langle \xi \vert X_{\gamma} Z_{\mu} [\sum_{\nu,\eta} \frac{1}{2^{2N}} \chi(\alpha \nu + \beta \eta)[\langle \xi \vert Z_{\nu} X_{\eta} \vert \xi \rangle ]^{-1}Z_{\nu}X_{\eta}]Z_{\alpha'} X_{\beta'}\vert \xi\rangle
\end{aligned}}
\end{equation*}
\begin{equation*}
\mathfit{97}{\begin{aligned}
& \times (3^{-1/4})^{h(\alpha'+\mu)+h(\beta'+\gamma)+h(\alpha'+\mu+\beta'+\gamma)} \\
& \times(\mathit{i}^{1/2})^{h(\alpha'+\mu)+h(\beta'+\gamma)-h(\alpha'+\mu+\beta'+\gamma)} \\
& \times \chi(\alpha'\beta' +\beta'\mu) \delta_{h(\alpha),m}\delta_{h(\beta),n}\delta_{h(\alpha+\beta),k}\delta_{h(\alpha'),m'}\\
& \times \delta_{h(\beta'),n'}  \delta_{h(\alpha'+\beta'),k'}\delta_{h(\mu),p}\delta_{h(\gamma),q}\delta_{h(\mu+\gamma),r},
\end{aligned}}
\end{equation*}
Expanding $[\langle\xi\vert Z_\nu X_\eta\vert\xi\rangle]^{-1}$ and applying the $Z$ and $X$ operators gives

\begin{equation*}
\mathfit{99}{\begin{aligned}
& =\sum_{\alpha,\beta,\alpha',\beta',\mu,\gamma}\langle \xi \vert X_{\gamma} Z_{\mu} [\sum_{\nu,\eta} \frac{1}{2^{2N}} \chi(\alpha \nu + \beta \eta)\\
& \times [ (3^{1/4})^{h(\nu)+h(\eta)+h(\nu+\eta)}(\mathit{i}^{-1/2})^{h(\nu)+h(\eta)-h(\nu+\eta)}]\\
& \times Z_{\nu}X_{\eta}]Z_{\alpha'} X_{\beta'}\vert \xi\rangle (3^{-1/4})^{h(\alpha'+\mu)+h(\beta'+\gamma)+h(\alpha'+\mu+\beta'+\gamma)} \\
& \times (\mathit{i}^{1/2})^{h(\alpha'+\mu)+h(\beta'+\gamma)-h(\alpha'+\mu+\beta'+\gamma)}\\
& \times\chi(\alpha'\beta'+\beta'\mu) \delta_{h(\alpha),m}\delta_{h(\beta),n}\delta_{h(\alpha+\beta),k} \\
& \times \delta_{h(\alpha'),m'}\delta_{h(\beta'),n'}\delta_{h(\alpha'+\beta'),k'}\delta_{h(\mu),p}\delta_{h(\gamma),q} \delta_{h(\mu+\gamma),r}
\end{aligned}}
\end{equation*}
\begin{equation*}
\mathfit{102}{\begin{aligned}
& = \sum_{\substack{\alpha,\beta,\alpha',\beta'\\ \mu,\gamma}}[\sum_{\nu,\eta} \frac{1}{2^{2N}} \chi(\alpha \nu + \beta \eta)[ (3^{1/4})^{h(\nu)+h(\eta)+h(\nu+\eta)}\\
& \times (\mathit{i}^{-1/2})^{h(\nu)+h(\eta)-h(\nu+\eta)} ]] \langle \xi \vert Z_{\alpha'+\mu+\nu} X_{\beta'+\eta+\gamma} \vert \xi\rangle \\
& \times  \chi(\gamma \mu+\gamma \nu) \chi(\alpha'\gamma+\alpha'\eta) \\
& \times (3^{-1/4})^{h(\alpha'+\mu)+h(\beta'+\gamma)+h(\alpha'+\mu+\beta'+\gamma)} \\
&\times (\mathit{i}^{1/2})^{h(\alpha'+\mu)+h(\beta'+\gamma)-h(\alpha'+\mu+\beta'+\gamma)}\\
& \times\chi(\alpha'\beta'+\beta'\mu) \delta_{h(\alpha),m}\delta_{h(\beta),n}\delta_{h(\alpha+\beta),k} \\
&\times \delta_{h(\alpha'),m'}\delta_{h(\beta'),n'}\delta_{h(\alpha'+\beta'),k'}\delta_{h(\mu),p}\delta_{h(\gamma),q} \delta_{h(\mu+\gamma),r}
\end{aligned}}
\end{equation*}
\begin{equation*}
\mathfit{105}{\begin{aligned}
& =\frac{1}{2^{2N}} \sum_{\substack{\alpha, \beta, \alpha', \beta'\\ \mu, \gamma, \nu, \eta}} \chi(\alpha \eta+\beta \eta) (3^{1/4})^{h(\nu)+h(\eta)+h(\nu+\eta)}\\
& \times (\mathit{i}^{-1/2})^{h(\nu)+h(\eta)-h(\nu+\eta)}\chi(\gamma \mu +\gamma \nu) \chi(\alpha'\gamma+\alpha'\eta)\\ 
& \times (3^{-1/4})^{h(\alpha'+\mu+\nu)+h(\beta'+\gamma+\eta)+h(\alpha'+\mu+\nu+\beta'+\gamma+\eta)} \\
& \times (i^{1/2})^{h(\alpha'+\mu+\nu)+h(\beta'+\gamma+\eta)-h(\alpha'+\mu+\nu+\beta'+\gamma+\eta)} \\
& \times (3^{-1/4})^{h(\alpha'+\mu)+h(\beta'+\gamma)+h(\alpha'+\mu+\beta'+\gamma)} \\
& \times (\mathit{i}^{1/2})^{h(\alpha'+\mu)+h(\beta'+\gamma)-h(\alpha'+\mu+\beta'+\gamma)} \\
& \times\chi(\alpha'\beta'+\beta'\mu) \delta_{h(\alpha),m}\delta_{h(\beta),n}\delta_{h(\alpha+\beta),k} \\
& \times \delta_{h(\alpha'),m'}\delta_{h(\beta'),n'}\delta_{h(\alpha'+\beta'),k'}\delta_{h(\mu),p}\delta_{h(\gamma),q} \delta_{h(\mu+\gamma),r}.
\end{aligned}}
\end{equation*}
To continue, the character has the alternative form \cite{CIS}

\begin{equation*}
\mathfit{109}{\begin{aligned}
& \chi(\alpha\beta) = (-1)^{\alpha \beta},
\end{aligned}}
\end{equation*}
The weights $h(\alpha)$, $h(\beta)$, and $h(\alpha+\beta)$ are sums of the nonzero components of the finite-field elements in the basis $\lbrace\theta_i\rbrace$ \cite{SymmetricDCS}.

Finally, using the Kronecker-delta representation $\delta_{\alpha,\beta}=(2\pi)^{-1}\int_0^{2\pi}e^{i(\alpha-\beta)\phi}\,d\phi$, the preceding expression can be written as

\begin{equation*}
\mathfit{110}{\begin{aligned}
& =\frac{1}{2^{2N}} \sum_{\substack{\alpha_i, \beta_i, \alpha_i', \beta_i' \\ \mu_i, \gamma_i, \nu_i, \eta_i}}(-1)^{\substack{\alpha_i \eta_i +\beta_i \mu_i + \gamma_i \mu_i + \gamma_i \nu_i +\alpha_i' \gamma_i +\alpha_i' \eta_i + \alpha_i' \beta_i' + \beta_i' \mu_i}}\\
& \times (3^{1/2})^{\nu_i + \eta_i - \nu_i \eta_i}  (3^{-1/2})^{\substack{\alpha_i' + \mu_i + \nu_i + \beta_i' + \gamma_i + \eta_i - \alpha_i' \mu_i \nu_i - \beta_i' \gamma_i \eta_i}}\\ 
& \times (3^{-1/2})^{\alpha_i '+\mu_i + \beta_i' + \gamma_i - \alpha_i' \mu_i -\beta_i' \gamma_i} (i)^{\nu_i \eta_i} (i)^{-\alpha_i' \mu_i \nu_i - \beta_i' \gamma_i \eta_i} (i)^{ - \alpha_i' \mu_i -\beta_i' \gamma_i}\\
& \times  \frac{1}{(2 \pi)^{9} } \Int_{0}^{2 \pi} \Int_{0}^{2 \pi} \Int_{0}^{2 \pi}  e^{-\mathit{i}xm+\mathit{i}x \alpha_i} e^{-\mathit{i}zn+\mathit{i}z \beta_i} e^{-\mathit{i}yk+\mathit{i}y(\alpha_i +\beta_i -2\alpha_i\beta_i) } dx dy dz \\
& \times \Int_{0}^{2 \pi} \Int_{0}^{2 \pi} \Int_{0}^{2 \pi}  e^{-\mathit{i}x'm'+\mathit{i}x' \alpha_i'} e^{-\mathit{i}z'n'+\mathit{i}z' \beta_i'} e^{-\mathit{i}y'k'+\mathit{i}y'(\alpha_i' +\beta_i' -2\alpha_i'\beta_i') }  dx' dy' dz' \\
& \times \Int_{0}^{2 \pi} \Int_{0}^{2 \pi} \Int_{0}^{2 \pi}  e^{-\mathit{i}up+\mathit{i}u \mu_i} e^{-\mathit{i}wq+\mathit{i}w \gamma_i} e^{-\mathit{i}vr+\mathit{i}v(\mu_i +\gamma_i -2\mu_i\gamma_i) } du dv dw,
\end{aligned}}
\end{equation*}
Grouping terms gives

\begin{equation*}
\mathfit{115}{\begin{aligned}
& =\frac{1}{2^{2N}} \frac{1}{(2 \pi)^{9} } \sum_{\substack{\alpha_i, \beta_i, \alpha_i', \beta_i' \\ \mu_i, \gamma_i, \nu_i, \eta_i}} (-1)^{\substack{\alpha_i \eta_i +\beta_i \mu_i + \gamma_i \mu_i + \gamma_i \nu_i +\alpha_i' \gamma_i +\alpha_i' \eta_i + \alpha_i' \beta_i' + \beta_i' \mu_i}} \\
&\times  (3^{-1/2})^{-(2 \alpha_i' + 2 \beta_i' + 2 \mu_i + 2 \gamma_i)} (3^{-1/2})^{\nu_i \eta_i -\alpha_i'\mu_i - \beta_i' \gamma_i }\\
& \times (3^{-1/2})^{- \alpha_i' \mu_i \nu_i - \beta_i' \gamma_i \eta_i} (i)^{-\alpha_i' \mu_i \nu_i - \beta_i' \gamma_i \eta_i} (i)^{ \nu_i \eta_i - \alpha_i' \mu_i -\beta_i' \gamma_i} \\
& \times  \Int_{0}^{2 \pi} \Int_{0}^{2 \pi} \Int_{0}^{2 \pi}  e^{-\mathit{i}xm+\mathit{i}x \alpha_i} e^{-\mathit{i}zn+\mathit{i}z \beta_i} e^{-\mathit{i}yk+\mathit{i}y(\alpha_i +\beta_i -2\alpha_i\beta_i) } dx dy dz \\
& \times \Int_{0}^{2 \pi} \Int_{0}^{2 \pi} \Int_{0}^{2 \pi}  e^{-\mathit{i}x'm'+\mathit{i}x' \alpha_i'} e^{-\mathit{i}z'n'+\mathit{i}z' \beta_i'} e^{-\mathit{i}y'k'+\mathit{i}y'(\alpha_i' +\beta_i' -2\alpha_i'\beta_i') } dx' dy' dz' \\
& \times \Int_{0}^{2 \pi} \Int_{0}^{2 \pi} \Int_{0}^{2 \pi} e^{-\mathit{i}up+\mathit{i}u \mu_i} e^{-\mathit{i}wq+\mathit{i}w \gamma_i} e^{-\mathit{i}vr+\mathit{i}v(\mu_i +\gamma_i -2\mu_i\gamma_i) } du dv dw,
\end{aligned}}
\end{equation*}
Finally, grouping within the sum all terms that depend on $(\alpha_i,\beta_i,\alpha_i',\beta_i',\mu_i,\gamma_i,\nu_i,\eta_i)$ yields

\begin{equation}
\mathfit{120}{\begin{aligned}
& \Psi^{m,n,k,m',n',k'}_{p,q,r}=\frac{1}{2^{2N}}\frac{1}{(2 \pi)^{9} }\Int_{0}^{2 \pi} \Int_{0}^{2 \pi} \Int_{0}^{2 \pi} \Int_{0}^{2 \pi} \Int_{0}^{2 \pi} \Int_{0}^{2 \pi} \Int_{0}^{2 \pi} \Int_{0}^{2 \pi} \Int_{0}^{2 \pi}\\
& \times e^{\substack{-\mathit{i}xm - \mathit{i}zn-\mathit{i}yk - \mathit{i}x'm' - \mathit{i}z'n' - \mathit{i}y'k' - \mathit{i}up - \mathit{i}wk - \mathit{i}vq}} \\
& \times \sum_{\substack{\alpha_i, \beta_i, \alpha_i', \beta_i' \\ \mu_i, \gamma_i, \nu_i, \eta_i}} e^{\substack{\mathit{i}x \alpha_i+\mathit{i}z \beta_i \mathit{i}y (\alpha_i +\beta_i -2\alpha_i \beta_i)+\mathit{i}x' \alpha_i'+\mathit{i}z' \beta_i'\mathit{i}y'\\ \times (\alpha_i'+\beta_i' -2\alpha_i'\beta_i') \mathit{i}u \mu_i+i w \gamma_i + \mathit{i}v(\mu_i +\gamma_i -2\mu_i\gamma_i)}}\\
& \times (3^{-1/2})^{\substack{\nu_i\eta_i-(2 \alpha_i' + 2 \beta_i' + 2 \mu_i + 2 \gamma_i) +(\alpha_i'\mu_i - \beta_i' \gamma_i )-(\alpha_i' \mu_i \nu_i +\beta_i' \gamma_i \eta_i)}} \\
& \times (i)^{\substack{\nu_i \eta_i-(\alpha_i' \mu_i \nu_i + \beta_i' \gamma_i \eta_i) +( \alpha_i' \mu_i -\beta_i' \gamma_i)}}\\
& \times (-1)^{\substack{\alpha_i \eta_i +\beta_i \mu_i + \gamma_i \mu_i + \gamma_i \nu_i +\alpha_i' \gamma_i +\alpha_i' \eta_i + \alpha_i' \beta_i' + \beta_i' \mu_i}}\\
& \times dx dy dz dx' dy' dz' du dv dw.
\end{aligned}}\label{eq:A_3}
\end{equation}
\section{The polynomial}
As described in the methodology, evaluating the sums in the expression for $\Psi^{m,n,k,m',n',k'}_{p,q,r}$, Eq.~(A.3), gives the following polynomial:

\begin{equation*}
\mathfit{124}{\begin{aligned}
& [ 4-\frac{2}{3}\,e^{ip}e^{iq}e^{i{\it n'}}e^{i{\it k'}}-\frac{2}{3}\,e^{ip}e^{ir}e^{in}e^{ik}-\frac{2}{3} e^{ip}e^{iq}e^{i{\it m'}}e^{i{\it k'}} \\
& -\frac{2}{3}\,e^{im}e^{ik}e^{i{\it n'}}e^{i{\it k'}}-\frac{2}{3}\,e^{im}e^{in}e^{i{\it n'}}e^{i{\it k'}} -\frac{2}{3}\,e^{ip}e^{ir}e^{i{\it m'}}e^{i{\it n'}} \\
&-\frac{2}{3}\,e^{iq}e^{ir}e^{im}e^{in} -\frac{2}{3} e^{iq}e^{ir}e^{im}e^{ik} -\frac{2}{3}\,e^{iq}e^{ir}e^{i{\it m'}}e^{i{\it n'}}\\
& +\frac{4}{3}\,e^{ip}e^{ir}e^{im}e^{ik} +\frac{4}{3}\,e^{im}e^{ik}e^{i{\it m'}}e^{i{\it k'}}-\frac{2}{3}\,e^{in}e^{ik}e^{i{\it m'}}e^{i{\it k'}} \\
& -\frac{2}{3}\,e^{im}e^{in}e^{i{\it m'}}e^{i{\it k'}}-\frac{2}{3}\,e^{ip}e^{ir}e^{i{\it n'}}e^{i{\it k'}}  -\frac{2}{3}\,e^{ip}e^{iq}e^{in}e^{ik}
\end{aligned}}
\end{equation*}
\begin{equation*}
\mathfit{127}{\begin{aligned}
& -\frac{2}{3}\,e^{iq}e^{ir}e^{i{\it m'}}e^{i{\it k'}}-\frac{2}{3}\,e^{im}e^{ik}e^{i{\it m'}}e^{i{\it n'}} -\frac{2}{3}\,e^{in}e^{ik}e^{i{\it m'}}e^{i{\it n'}}\\
& +\frac{4}{3}\,e^{im}e^{in}e^{i{\it m'}}e^{i{\it n'}} +\frac{4}{3}\,e^{ip}e^{iq}e^{im}e^{in} -\frac{2}{3} e^{ip}e^{iq}e^{im}e^{ik}\\
& -\frac{2}{3} e^{ip}e^{ir}e^{im}e^{in}+\frac{4}{3} e^{in} e^{ik}e^{i n'} e^{i k'} +\frac{4}{3} e^{iq} e^{ir} e^{in} e^{ik}\\
& +\frac{4}{3} e^{ip} e^{iq}+\frac{4}{3} e^{im'}e^{i k'}+\frac{4}{3}\,e^{ip}e^{ir}+\frac{4}{3}\,e^{i{\it n'}}e^{i{\it k'}}+\frac{4}{3}\,e^{i{\it m'}}e^{i{\it n'}} \\
& +\frac{2 i}{\sqrt{3}}\,e^{iq}e^{ir}e^{im}e^{ik}e^{i{\it m'}}e^{i{\it n'}}-\frac{2 i}{\sqrt{3}} e^{ip}e^{iq}e^{im}e^{ik}e^{i{\it n'}}e^{i{\it k'}}\\
& -\frac{2 i}{\sqrt{3}}\,e^{ip}e^{ir}e^{in}e^{ik}e^{i{\it m'}}e^{i{\it n'}}+\frac{2 i}{\sqrt{3}} e^{ip}e^{iq}e^{in}e^{ik}e^{i{\it m'}}e^{i{\it k'}} \\
& -\frac{2 i}{\sqrt{3}}\,e^{iq}e^{ir}e^{im}e^{in}e^{i{\it m'}}e^{i{\it k'}}+\frac{4}{3}\,e^{iq}e^{ir}\\
& +\frac{4}{3}\,e^{iq}e^{ir}e^{in}e^{ik}e^{i{\it m'}}e^{i{\it k'}}-\frac{2}{3}\,e^{iq}e^{ir}e^{im} e^{in}e^{i{\it m'}}e^{i{\it k'}} \\
& -\frac{2}{3}\,e^{iq}e^{ir}e^{im}e^{ik}e^{i{\it m'}}e^{i{\it n'}}+\frac{4}{3}\,e^{ip}e^{iq}e^{in} e^{ik}e^{i{\it n'}}e^{i{\it k'}}\\
& +4\,e^{ip}e^{iq}e^{im}e^{in}e^{i{\it m'}}e^{i{\it n'}}+\frac{4}{3}\,e^{ip}e^{ir}e^{im}e^{ik}e^{i{\it m'}} e^{i{\it n'}} \\
& +\frac{4}{3}\,e^{ip}e^{iq}e^{im}e^{ik}e^{i{\it m'}}e^{i{\it k'}}-\frac{2}{3}\,e^{ip}e^{iq}e^{in} e^{ik}e^{i{\it m'}}e^{i{\it k'}}\\
& +\frac{4}{3}\,e^{ip}e^{iq}e^{im}e^{in}e^{i{\it m'}}e^{i{\it k'}}-\frac{2}{3}\,e^{ip}e^{ir}e^{in} e^{ik}e^{i{\it m'}}e^{i{\it n'}} \\
& +\frac{4}{3}\,e^{ip}e^{ir}e^{im}e^{in}e^{i{\it m'}}e^{i{\it n'}}+\frac{4}{3}\,e^{ip}e^{iq}e^{im} e^{in}e^{i{\it n'}}e^{i{\it k'}}\\
& -\frac{2}{3}\,e^{ip}e^{iq}e^{im}e^{ik}e^{i{\it n'}}e^{i{\it k'}}+\frac{4}{3}\,e^{iq}e^{ir}e^{im} e^{in}e^{i{\it m'}}e^{i{\it n'}} \\
& +\frac{4}{3}\,e^{iq}e^{ir}e^{in}e^{ik}e^{i{\it m'}}e^{i{\it n'}}+\frac{4}{3}\,e^{ip}e^{ir}e^{im} e^{ik}e^{i{\it n'}}e^{i{\it k'}}\\
& +\frac{4}{3}\,e^{ip}e^{ir}e^{in}e^{ik}e^{i{\it n'}}e^{i{\it k'}}+4\,e^{iq}e^{ir}e^{in}e^{ik}e^{i{\it n'}} e^{i{\it k'}} \\
& + 4\,e^{ip}e^{ir}e^{im}e^{ik}e^{i{\it m'}}e^{i{\it k'}}+\frac{2 i}{\sqrt{3}}\,e^{ip}e^{iq}e^{i{\it n'}} e^{i{\it k'}}\\
& +\frac{2 i}{\sqrt{3}}\,e^{ip}e^{ir}e^{im}e^{in}e^{i{\it n'}}e^{i{\it k'}}+\frac{2 i}{\sqrt{3}} e^{ip}e^{ir}e^{i{\it m'}}e^{i{\it n'}} \\
& +\frac{2 i}{\sqrt{3}}\,e^{iq}e^{ir}e^{im}e^{in}-\frac{2 i}{\sqrt{3}}\,e^{iq}e^{ir}e^{im}e^{ik}\\
& +\frac{2 i}{\sqrt{3}}\,e^{ip}e^{iq}e^{im}e^{ik}+\frac{2 i}{\sqrt{3}}\,e^{in}e^{ik}e^{i{\it m'}} e^{i{\it n'}} \\
& +\frac{2 i}{\sqrt{3}}\,e^{iq}e^{ir}e^{i{\it m'}}e^{i{\it k'}}-\frac{2 i}{\sqrt{3}}\,e^{im}e^{ik} e^{i{\it m'}}e^{i{\it n'}}\\
& -\frac{2 i}{\sqrt{3}}\,e^{ip}e^{iq}e^{in}e^{ik}-\frac{2 i}{\sqrt{3}}\,e^{ip}e^{ir}e^{i{\it n'}} e^{i{\it k'}} \\
& -\frac{2}{3}\,e^{ip}e^{ir}e^{im}e^{in}e^{i{\it n'}}e^{i{\it k'}}+\frac{4}{3}\,e^{iq}e^{ir}e^{im} e^{ik}e^{i{\it m'}}e^{i{\it k'}}\\
& -\frac{2 i}{\sqrt{3}}\,e^{iq}e^{ir}e^{i{\it m'}}e^{i{\it n'}}+\frac{2 i}{\sqrt{3}}\,e^{ip}e^{ir} e^{in}e^{ik} \\
& -\frac{2 i}{\sqrt{3}}\,e^{ip}e^{ir}e^{im}e^{in}-\frac{2 i}{\sqrt{3}}\,e^{in}e^{ik}e^{i{\it m'}} e^{i{\it k'}}\\
& +\frac{2 i}{\sqrt{3}}\,e^{im}e^{in}e^{i{\it m'}}e^{i{\it k'}}
-\frac{2 i}{\sqrt{3}}\,e^{ip}e^{iq}  e^{i{\it m'}}e^{i{\it k'}} \\
& +\frac{2 i}{\sqrt{3}}\,e^{im}e^{ik}e^{i{\it n'}}e^{i{\it k'}}-\frac{2 i}{\sqrt{3}}\,e^{im}e^{in}  e^{i{\it n'}}e^{i{\it k'}} ]^N
\end{aligned}}
\end{equation*}

\end{document}